\documentstyle[dina4,epsfig,here]{article}
\begin{document}
\topmargin -3.cm
\textwidth 15.cm
\textheight 23.cm
\title{Production of Charged Pions, Kaons and Antikaons\\
in Relativistic C+C and C+Au Collisions}

\author{
F.~Laue (1)
I.~B\"ottcher (3)
M.~D\c{e}bowski (5)
A.~F\"orster (2)
E.~Grosse (6,7)\\
P.~Koczo\'n (1)
B.~Kohlmeyer (3)
M.~Mang (1)
M.~Menzel (3)
L.~Naumann (6)\\
H.~Oeschler (2)
F.~P\"uhlhofer (3)
E.~Schwab (1)
P.~Senger (1)
Y.~Shin (4)\\
J.~Speer (3)
H.~Str\"obele (4)
C.~Sturm (2)
G.~Sur\'owka (1,5)
F.~Uhlig (2)\\
A.~Wagner (6)
W.~Walu\'s (5)\\
(KaoS Collaboration)\\
(1) Gesellschaft f\"ur Schwerionenforschung, D-64220 Darmstadt, Germany\\
(2) Technische Universit\"at Darmstadt, D-64289 Darmstadt, Germany\\
(3) Phillips Universit\"at, D-35037  Marburg, Germany\\
(4) Johann Wolfgang Goethe Universit\"at, D-60325 Frankfurt am Main, Germany\\
(5) Jagiellonian University, PL-30059 Krak\'ow, Poland\\
(6) Forschungszentrum Rossendorf, D-01314 Dresden, Germany\\ 
(7) Technische Universit\"at Dresden, D-01314 Dresden, Germany
}
\maketitle


\begin{abstract}
Production cross sections of charged pions, 
kaons and antikaons  have been measured in C+C and C+Au collisions 
at beam energies of 1.0 and 1.8 AGeV for different polar emission angles.
The kaon and antikaon energy spectra can be described by Boltzmann distributions
whereas the pion spectra exhibit an additional enhancement at low energies.
The pion multiplicity per participating nucleon M($\pi^+$)/$<A_{part}>$ 
is a factor of about 3 smaller in C+Au  than in C+C collisions at 1.0 AGeV
whereas it differs only little for the C and the 
Au target at a beam energy of 1.8 AGeV. 
The K$^+$ multiplicities per participating nucleon M(K$^+$)/$<A_{part}>$ 
are independent of the target size at 1 AGeV and at 1.8 AGeV. 
The K$^-$ multiplicity per participating nucleon 
M(K$^-$)/$<A_{part}>$ is reduced by 
a factor of about 2  in C+Au as compared to C+C collisions at 1.8 AGeV. 
This effect might be caused by the absorption of antikaons 
in the heavy target nucleus.
Transport model calculations underestimate the K$^-$/K$^+$ ratio for 
C+C collisions at 1.8 AGeV  by a factor of about 4 if in-medium modifications 
of K mesons are neglected.      
\end{abstract}

\section{Introduction}
\label{intro}
The study of meson production and propagation in relativistic 
nucleus-nucleus collisions has become an important experimental 
tool for the investigation of the properties 
of nuclear matter at high densities and of the 
hadron self energy in a dense nuclear medium \cite{stock,cass_brat,sen_str}.  
The pion multiplicity measured in heavy-ion experiments at the BEVALAC
has been correlated with the 
thermal energy of the fireball
in order to extract information on the nuclear matter equation of 
state \cite{harris}. In a more advanced approach, the production of kaons
in nucleus-nucleus collisions at beam energies below the threshold
for free NN collisions (E$_{beam}$ = 1.58 GeV for NN$\to$K$^+\Lambda$N)
is used to probe  the compressibility of
nuclear matter \cite{aich_ko,li_ko}.

The properties of strange mesons in a medium of finite baryon density are
essential for our understanding of strong interactions.
According to various theoretical approaches, antikaons
feel strong attractive forces in the nuclear medium whereas
the in-medium kaon-nucleon potential is expected to be slightly repulsive
\cite{brown1,waas,schaffner,lutz}. Predictions
have been made that the effective
mass of the  K$^-$ meson decreases with increasing nuclear density
leading to K$^-$ condensation in neutron stars above 3
times saturation density $\rho_0$. This effect is
expected to influence significantly  the evolution of supernova explosions:
the K$^-$ condensate softens the nuclear equation of state
and thus causes a core with 1.5 - 2 solar masses to collapse
into a black hole rather than to form a neutron star \cite{brobet,li_lee_br}.

Heavy-ion  collisions at relativistic energies $-$
where baryonic densities of several times the saturation density
are reached $-$ offer the possibility to study in-medium properties of
strange mesons. In dense nuclear matter the 
K$^-$ effective mass is predicted to be reduced and thus
the kinematical threshold for the process NN$\to$K$^-$K$^+$NN
will be lowered. As a consequence, the
K$^-$ yield in nucleus-nucleus collisions at bombarding energies below the NN
threshold (E$_{beam}$ = 2.5 GeV for NN$\to$K$^+$K$^-$NN) 
is expected to be enhanced significantly as compared to the case
without in-medium mass reduction.
In contrast, the yield of K$^+$ mesons is predicted to be decreased
as the K$^+$ effective mass and thus the in-medium K$^+$ production
threshold is slightly increased \cite{cassing,li_ko_fang,schaffner}.

The large antikaon/kaon ratio measured in nucleus-nucleus
collisions at ''subthreshold'' beam energies 
has been interpreted as a signature for significant in-medium modifications 
of antikaons \cite{barth,laue,cassing,li_ko_fang}.
Moreover, anisotropies of the azimuthal
emission pattern of kaons have been measured and explained by an in-medium 
kaon nucleon potential \cite{shin,li_ko_br}. The same explanation was 
proposed for the observation of a vanishing in-plane 
flow of kaons \cite{ritman,li_flow}. 

The various ratios of particles produced in nucleus-nucleus collisions
can be used as a ''thermometer'' if an equilibrated fireball has been created
\cite{pbm1,cleymans}.
Whether this happens or not is still a matter of debate. 
Up to now, particle ratios measured in Ni+Ni and Au+Au at SIS energies
have been explained both by transport calculations and   
by  a thermal model which assumes a common 
freeze-out temperature \cite{cleymans}. 

In contrast to heavy collision systems, very light or strongly
asymmetric systems allow  to differentiate  
between theoretical descriptions such as thermal models or transport models. 
Predictions of statistical models $-$ which are based on the assumption 
of a hadronic fireball in thermal and chemical equilibrium $-$ 
might disagree with the data whereas 
transport models should be well suited 
for the description of nonequilibrated systems.
Moreover, compressional effects are negligible 
in light symmetric collision systems, 
and hence nearly no collective flow is created. 
Therefore, the particle spectra 
reflect only the ''thermal'' energy of the particles. In addition,
absorption effects should be small.  

Another important aspect of particle production in nucleus-nucleus
collisions is the absorption of particles in the cold spectator matter
or in the dilute stage of the collision. 
This question is crucial when  particle multiplicities 
are used as a ''calorimeter''  (to determine the thermal energy of a fireball)
or as a probe for in-medium effects (e.g. when particle yields
or ratios are related to effective masses in nuclear matter).  
This aspect can be studied in strongly asymmetric collisions where 
the large target spectator provides a source of particle absorption.  

Therefore, the comparison of particle yields in 
light symmetric  and strongly asymmetric
collision systems may shed light on 
(i) 
in-medium and absorption effects in the
absence of compression and flow
and on (ii) the predictive power of statistical and nonequilibrium  
models.     
There is also an experimental challenge connected with 
the interpretation of data measured in asymmetric collisions.
As the velocity of the particle emitting source is not known a priori
it is hard to disentangle the effects of the Lorentz-boost, 
the polar angle distribution and rescattering of particles from spectator 
matter.  These effects will be discussed below.  

Pioneering experiments on kaon production in nucleus-nucleus collisions
have been performed with a single arm spectrometer at the BEVALAC 
\cite{schnetzer}. Those measurements made use of the highest
possible beam energy of 2.1 AGeV 
in order to enhance the kaon-to-proton ratio. 
Antikaons have been measured in beam-line spectrometers at 
$\Theta_{lab}$ = 0$^{\circ}$  which  
have a small acceptance in solid angle and momentum but an efficient 
partice identification capability \cite{shor,schroeter}.
In this article we present experimental results on the production of pions,  
kaons and antikaons in C+C and 
C+Au collisions at beam energies of 1.0 and 1.8 AGeV 
measured at various laboratory angles. 
The paper is arranged as follows. In the next chapter the experimental setup
is described. Then the experimental results are presented: 
double-differential cross sections for the production of pions, kaons 
and antikaons in the laboratory frame for C+C and 
C+Au collisions and invariant cross sections for C+C collisions in the 
center-of-mass (c.m.) frame. For the symmetric system 
we find a nonisotropic polar angle distribution.   
Then we discuss the dependence of the particle multiplicities  
on the number of participating nucleons 
(A$_{part}$) in symmetric collisions and use the 
scaling behaviour to estimate A$_{part}$ for the asymmetric C+Au system.
We determine the velocity of the particle emitting source in C+Au collisions
with a transport calculation and discuss spectral shapes 
in the source frame.  
Finally we discuss in-medium effects and
compare the data to the predictions of transport calculations.

\section{Experimental setup}
The experiments reported here 
have been performed with the Kaon Spectrometer (KaoS) \cite{senger}
at the heavy-ion synchrotron SIS at GSI in Darmstadt.  
KaoS was designed to identify kaons and antikaons in collisions between
heavy nuclei at  very low beam energies: 
the  proton/pion/kaon ratio can be as large as 10$^6$/10$^4$/1.
In order to make use of the highest beam intensities 
the experiment is equipped with an efficient kaon trigger 
based on time-of-flight and Cherenkov detectors. 
Moreover, the KaoS setup is able to determine
the centrality of the collision, the number of participants and the 
orientation of the reaction plane.

\begin{figure*}[H]
\mbox{\epsfig{file=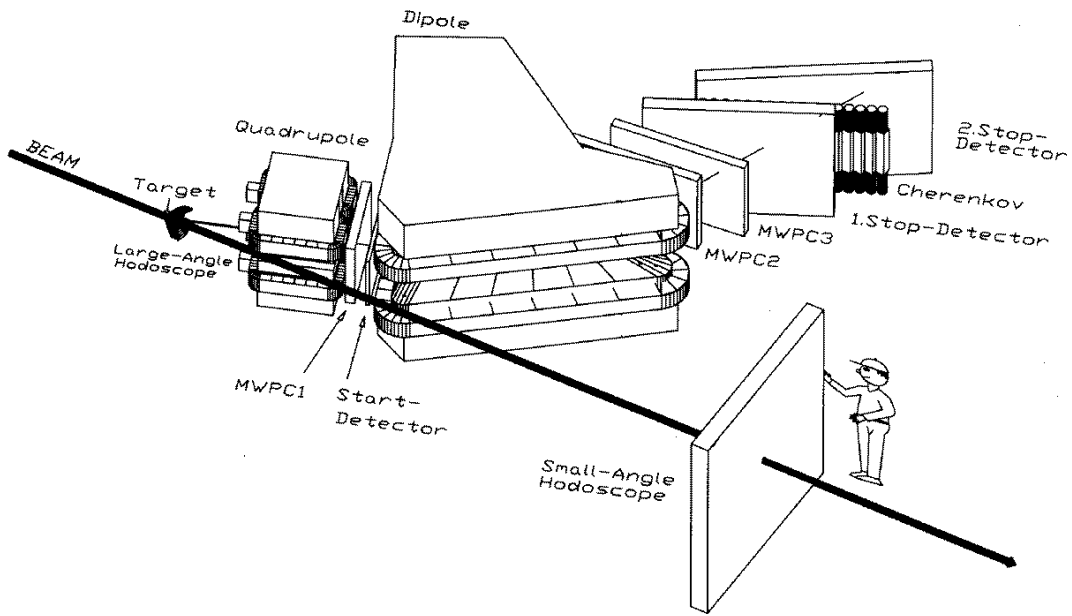,width=15.cm}}
\caption{Sketch of the Kaon Spectrometer and its detector system.
}
\label{kaos}
\end{figure*}

Figure \ref{kaos} shows a sketch of the experimental setup. 
The spectrometer consists of 
a double-focussing quadrupole-dipole combination with a large acceptance in 
momentum (p$_{max}$/p$_{min} \approx$ 2) and solid angle 
($\Omega \leq$ 35 msr).  The maximum  dipole magnetic field is  B = 1.95 T 
corresponding to a momentum  of p$_{max}$ = 1.6 GeV/c for  
particles with charge one.

The spectrometer is equipped with plastic scintillator arrays 
for time-of-flight measurements. 
The start detector  consists of 16 elements (length 220 mm, 
width 30 mm, thickness 4 mm) and  
is located between the quadrupole and the dipole magnet. The time 
resolution is about 320 ps FWHM. The stop detector consists of 30 elements 
(length 380 mm, width 37 mm, thickness 20 mm) and is positioned along 
the focal plane of the spectrometer. The time resolution is about 100 ps FWHM, 
the distance from start to stop detector is 3-4 m depending on trajectory. 

A plastic scintillator hodoscope (Large Angle Hodoscope, ''LAH'')
consisting of 84 modules is located 8 - 13 cm 
downstream of the target. The modules  are arranged in 3 rings and cover a
polar angle range of $ 12^{\circ} < \Theta_{lab} < 48^{\circ}$. This detector 
measures the multiplicity of charged particles which is correlated to the 
centrality of the collision. The time signals generated by the 
particles (mostly protons, deuterons and pions) in the LAH 
are used to determine  the time when the reaction takes place in the target
(t= 0). 
The time-of-flight  between LAH 
and the start detector (distance about 1.7 m) provides additional independent 
information  on the particle velocity. The comparison of the 
two measurements of the particle velocity (LAH-Start and Start-Stop) 
efficiently reduces the background of rescattered particles with 
false trajectories. 

The particle trajectories were measured with 
three large area multiwire proportional chambers (MWPC). The first MWPC has 
an active area of 30x60 cm$^2$ and is located between 
quadrupole and dipole. The 2. and 3. MWPC have
active areas of 35x120 cm$^2$ and are positioned between dipole magnet 
and stop detector. The position resolution of the chambers  is 1 mm 
FWHM. The quality of track recognition and vertex reconstruction 
is limited by small angle straggling in the detectors and in the air.  

A Cherenkov threshold detector is positioned behind the TOF stop wall 
\cite{misko2}.
The aim of this detector is to prevent high-energy protons from triggering 
the data aquisition, if the time-of-flight difference of protons and 
kaons is too small to be discriminated by the TOF trigger. The 
threshold velocity for particles producing Cherenkov light is 
$\beta$ = 1/n with n the refractive index of the radiator material. 
We chose lucite (n = 1.49) and water (n = 1.34) as radiators corresponding
to threshold velocities of $\beta$ = 0.67  and $\beta$ = 0.75, respectively.
The Cherenkov detector covers an area of 40x190 cm$^2$ and 
consists of 10 lucite modules at the low momentum side 
(height 40 cm, width 10 cm, thickness 5 cm) and 6 water modules at the high
momentum side (height 40 cm, width 10 cm, thickness 10 cm).
The overall efficiency of the Cherenkov detector for pions, kaons and protons
is 0.96, 0.86 and 0.02, respectively, at particle momenta between 
0.64 and 1.14 GeV/c. 

The (hardware) kaon trigger is based on TOF and Cherenkov information. 
The kaon TOF trigger  
rejects all particles which differ in time-of-flight by a few nanoseconds 
from the  nominal value of the kaon  time-of-flight 
for a given magnetic field configuration.   
Depending on the magnetic field of the spectrometer, the kaon TOF  
trigger reduces the proton trigger rate by a factor of 10 - 100 
and the pion trigger rate by a factor of about 10. 
At high magnetic fields where the TOF difference 
between protons and kaons becomes small, the requirement of a Cherenkov signal
reduces the proton trigger rate by another factor of about 10 (with the 
TOF trigger active).  

\begin{figure}[H]
\mbox{\epsfig{file=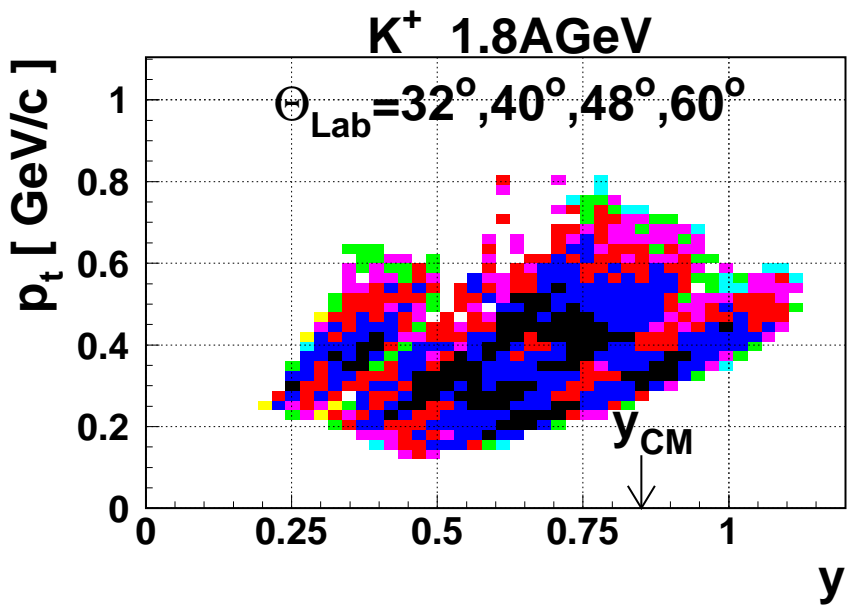,width=12.cm}}
\caption{ Phase space covered by the Kaon Spectrometer 
with 4 angles and 3 magnetic fields in the transverse-momentum
versus rapidity plane. The scatter plot of measured kaons is
not corrected for acceptance.   
}
\label{phasespace}
\end{figure}

The spectrometer can be pivoted on the target point  
within the angular range of $ 0^{\circ} < \Theta_{lab} < 130^{\circ}$. 
Fig.~\ref{phasespace} 
displays as an example the coverage of phase space achieved with 4 angles
and 3 magnetic fields for kaons measured at 1.8 AGeV beam energy. 

The beam intensity is determined with two plastic scintillator 
telescopes which consist of 3 detectors (40x40 mm$^2$) each.   
The telescope arms point to the target at $\Theta_{lab}$ = 100$^{\circ}$ in
the horizontal plane.
The distance between the target and the first detector and between the
subsequent detectors is 15 cm. The counting rate of the threefold coincidences 
of the detectors of each telescope arm is recorded with a scaler. 
This counting rate is proportional to the beam current. In fact, 
we take the average
counting rate of the two telescopes in order to correct for  a (horizontally) 
noncentral beam position at the target. The calibration factor 
(beam intensity/telescope counting rate) is determined by reducing the 
beam intensity to about 50000 ions/s and counting the beam directly with 
a movable plastic scintillator (20x20x3mm$^3$).

The challenge of the experiment is the identification of kaons and antikaons 
in the presence of a large background. This background is caused by 
(i) rescattered particles, mostly high-energetic 
protons which hit the yoke of the quadrupole and are deflected
into the focal plane detectors and (ii) two particles from different reactions
producing a random coincidence in the start/stop detectors
and spurious tracks in the MWPC's.
The reconstruction of these false tracks may result in an entry at a mass
around 500 MeV/c$^2$ which contributes to the ''background'' below 
the kaon mass peak.
Track recognition is performed by measuring hits in 
the second and third  MWPC. These tracks are extrapolated 
to the first MWPC in which the  calculated hit positions are compared to
the measured ones. The calculation of the trajectories is based on 
GEANT simulations. Moreover, the tracks are extrapolated to the 
start and stop detectors  and the calculated positions are compared to those 
paddles which had been hit by particles. Finally, the particle velocity
measured between the LAH  and the start detector is compared to the one 
measured between the start and the stop detector. A track is rejected, if 
the difference $\Delta$x (between measured and calculated positions) 
and the velocity difference $\Delta\beta$ (measured with the two 
TOF sections) exceed a value of 3 $\sigma$  of the $\Delta$x and 
$\Delta\beta$ distributions, respectively. The losses   
due to these cuts are determined by kaon data with low background, 
i.e. kaon measurements
at high bombarding energies (1.8-2.0 AGeV). These cuts are 
used also for the analysis of antikaons. 
The sum of all losses is about 20\%. 

The geometrical acceptance of the spectrometer and kaon decay in flight 
is determined by Monte Carlo simulations using the code GEANT. 
The kaon trigger efficiency is measured 
with so called ''pseudokaons'' which are protons (or pions)
having the  same velocity as the kaons in the experiment. This is achieved by 
choosing appropriate  magnetic fields of dipole and quadrupole 
and using the same TOF-trigger conditions as for the kaon measurement.          
Similarly, the efficiency of the Cherenkov trigger is determined. The kaon 
trigger
efficiency is about 90\%. The efficiency of the MWPC's for the identification
of minimum ionizing particles is measured to be better than  95\%.

\begin{figure*}[H]
\centerline{
\mbox{\epsfig{file=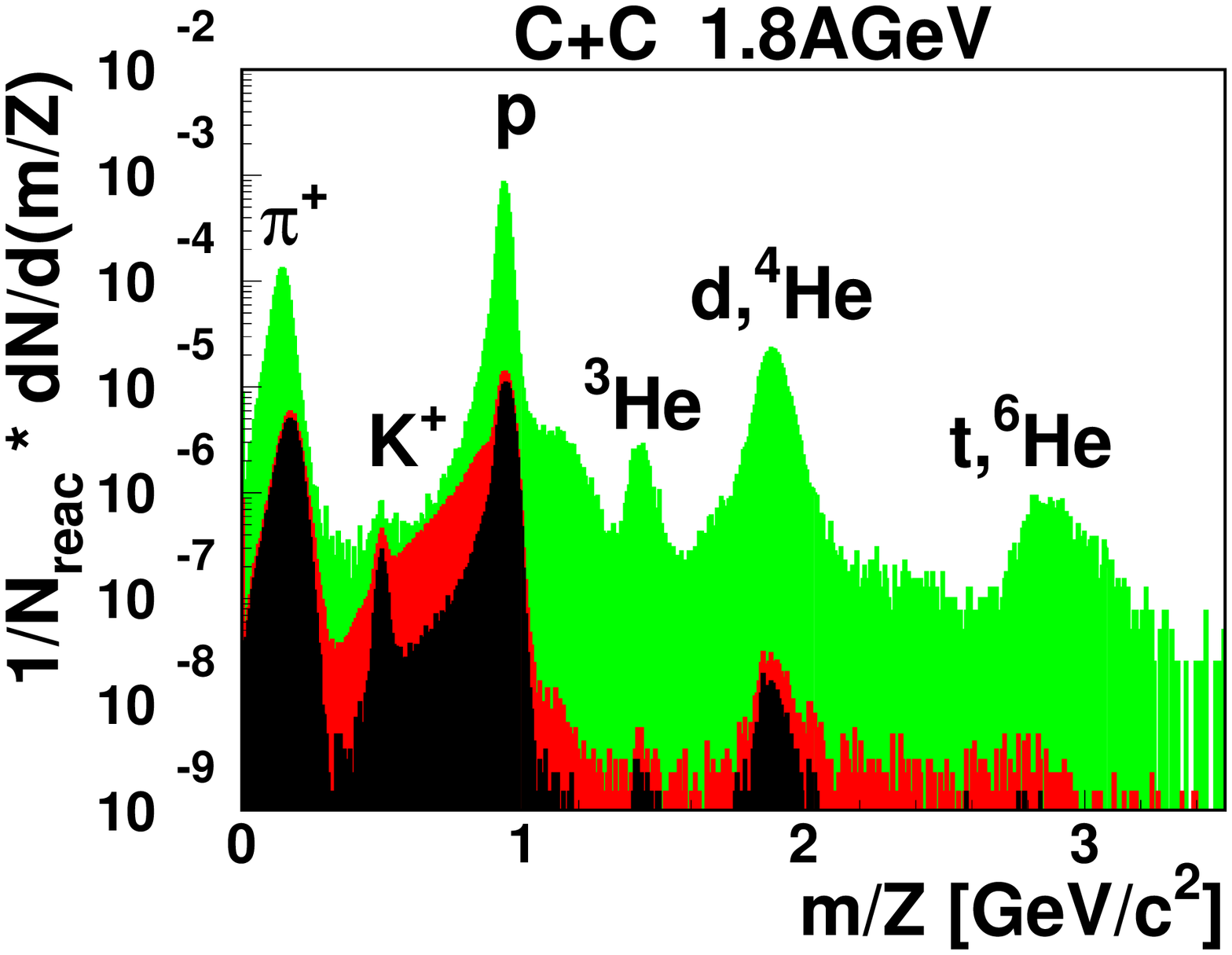,width=8.cm}}
\mbox{\epsfig{file=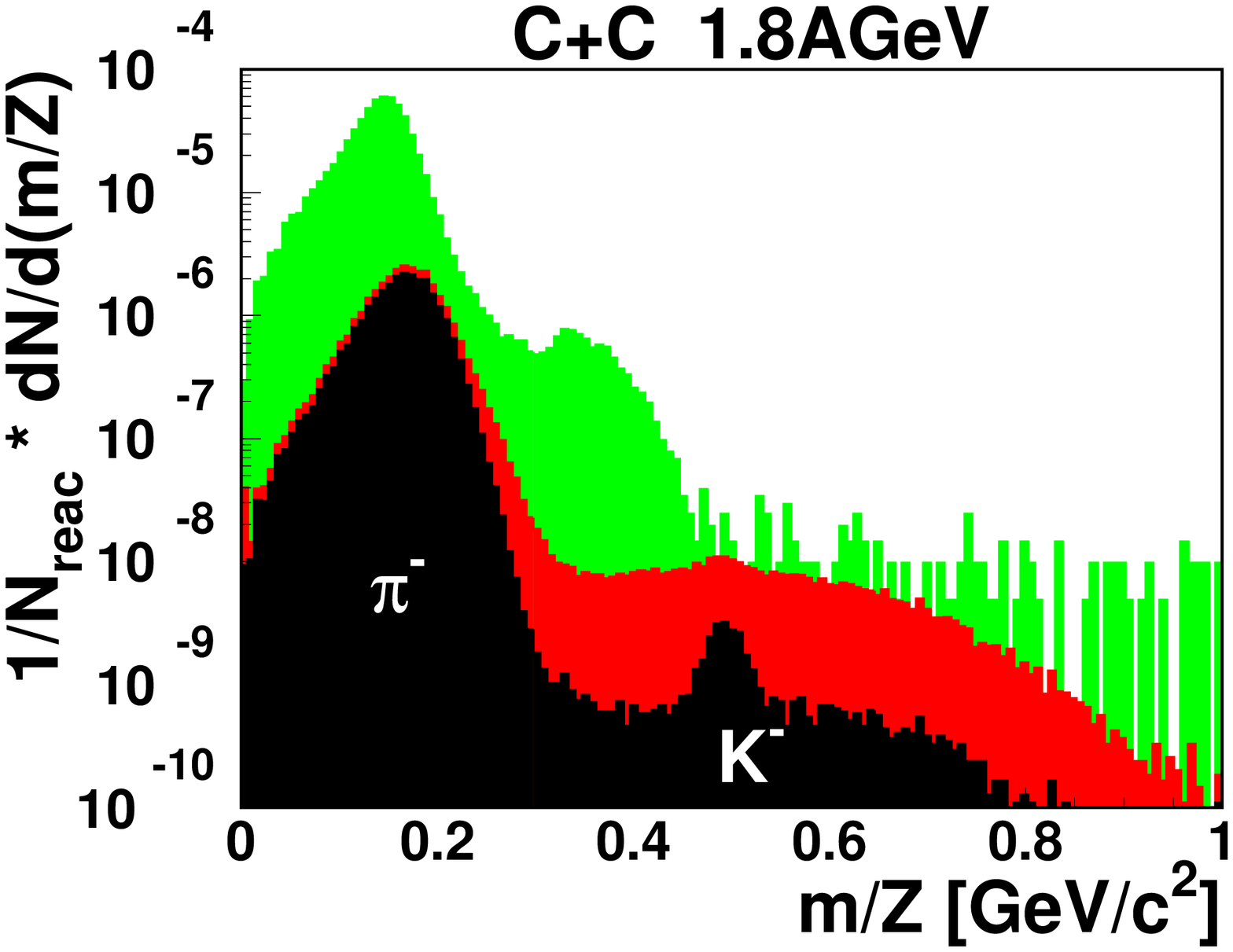,width=8.cm}}
}
\caption{ Mass/Z distribution of positively (left) and negatively (right)
charged particles for different  trigger and tracking conditions.
}
\label{mass}
\end{figure*}

The quality of particle identification in C+C collisions at 1.8 AGeV 
is shown in fig.~\ref{mass} for positively (left) and negatively 
charged particles (right) as  function
of their mass over charge ratio m/Z.  The light grey spectrum is measured
with a trigger on charged
particles in the spectrometer without time-of-flight condition.
The medium grey spectrum is measured  with the TOF trigger. The dark
spectrum is measured with the kaon trigger 
and analysed with conditions on tracking and particle velocity.  
The background near to the K mass is reduced by more than two orders 
of magnitude. What remains is quantified by a fit and is substracted.

The measurements reported here have been performed 
with a Carbon beam. The beam energy, the target isotope
and thickness, the measured kaon rates and numbers are listed in 
table \ref{table1}.

\section{Experimental results}

The systematic errors of the measured cross sections presented in this chapter
amount to about 15\%. These errors are due to the  uncertainties of 
beam normalization (7\%), efficiencies of the TOF-trigger (5\%), 
detector efficiencies  (2\%), track recognition
(7\%), kaon background subtraction (5\%), dead time correction (7\%) and 
acceptance calculation by GEANT (5\%).

\subsection{Differential production cross sections}

\begin{figure*}[H]
\centerline{
\mbox{\epsfig{file=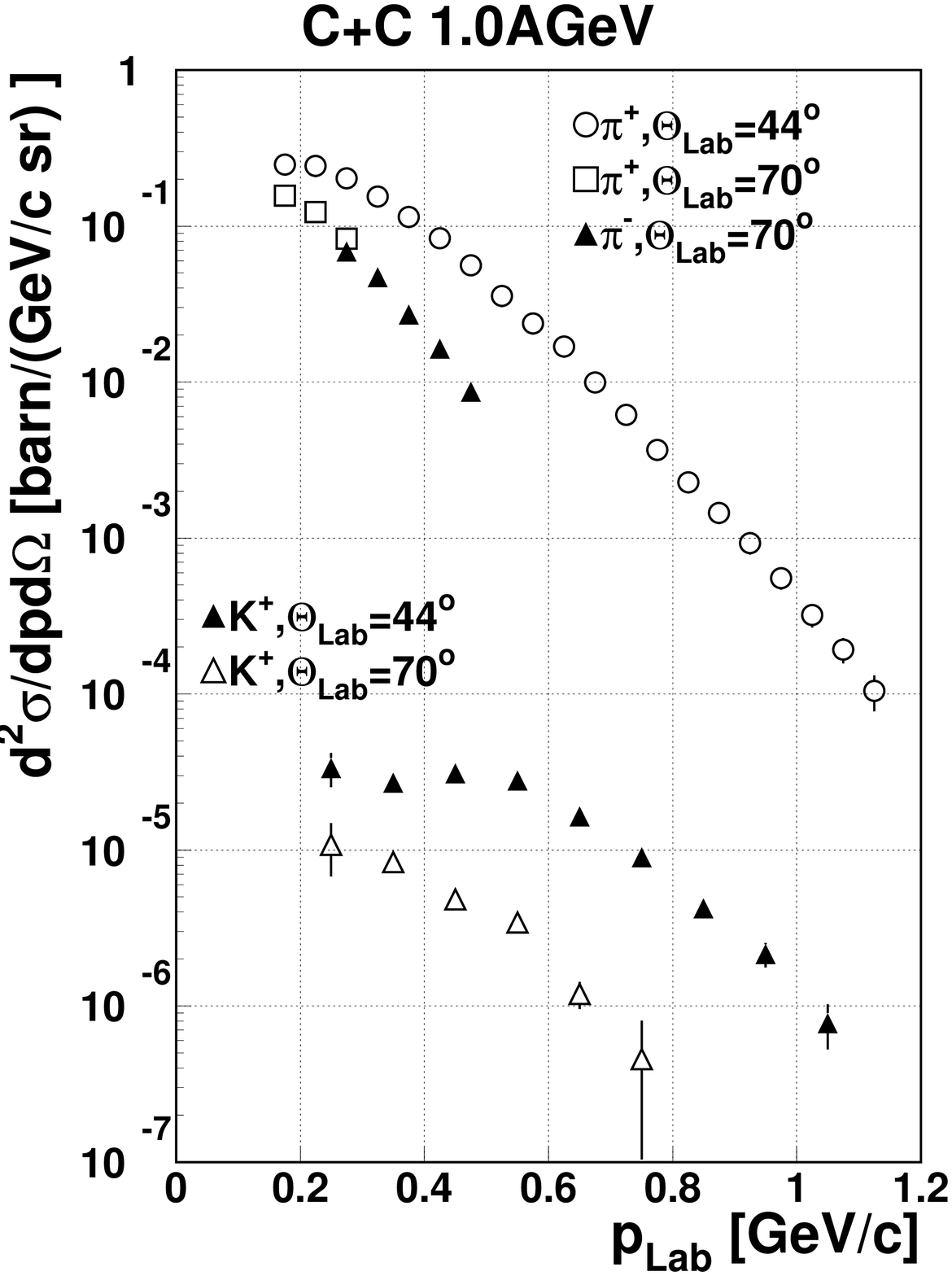,width=7.cm}}
\mbox{\epsfig{file=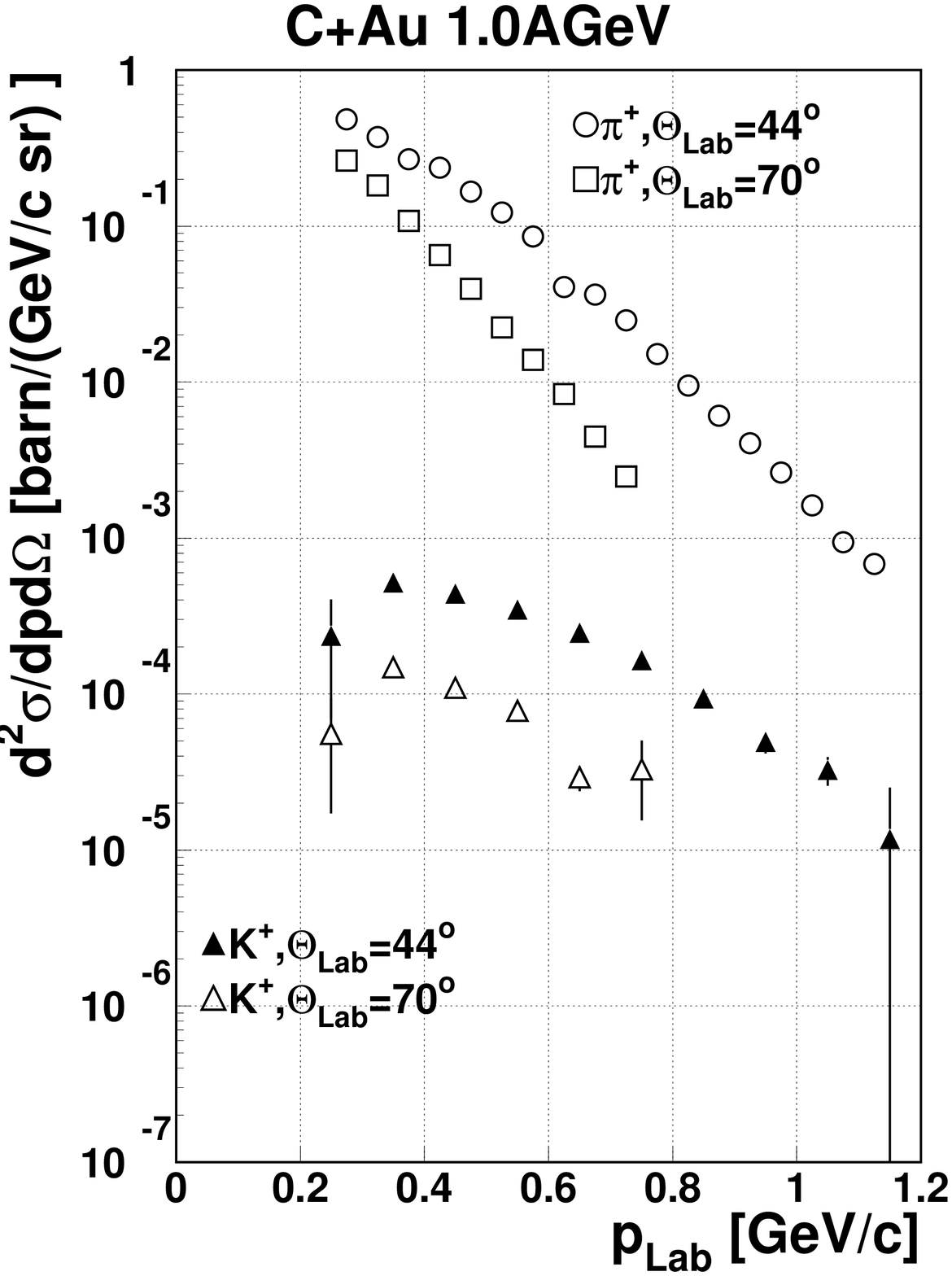,width=7.cm}}
}
\caption{ Inclusive double-differential cross-sections for the production of 
pions and K$^+$ mesons measured in C+C (left) and C+Au (right) 
collisions at a beam energy of 1.0 AGeV
under different  laboratory angles (as indicated) as function of laboratory
momentum.
}
\label{data_cc10_lab}
\end{figure*}

\begin{figure*}[H]
\centerline{
\mbox{\epsfig{file=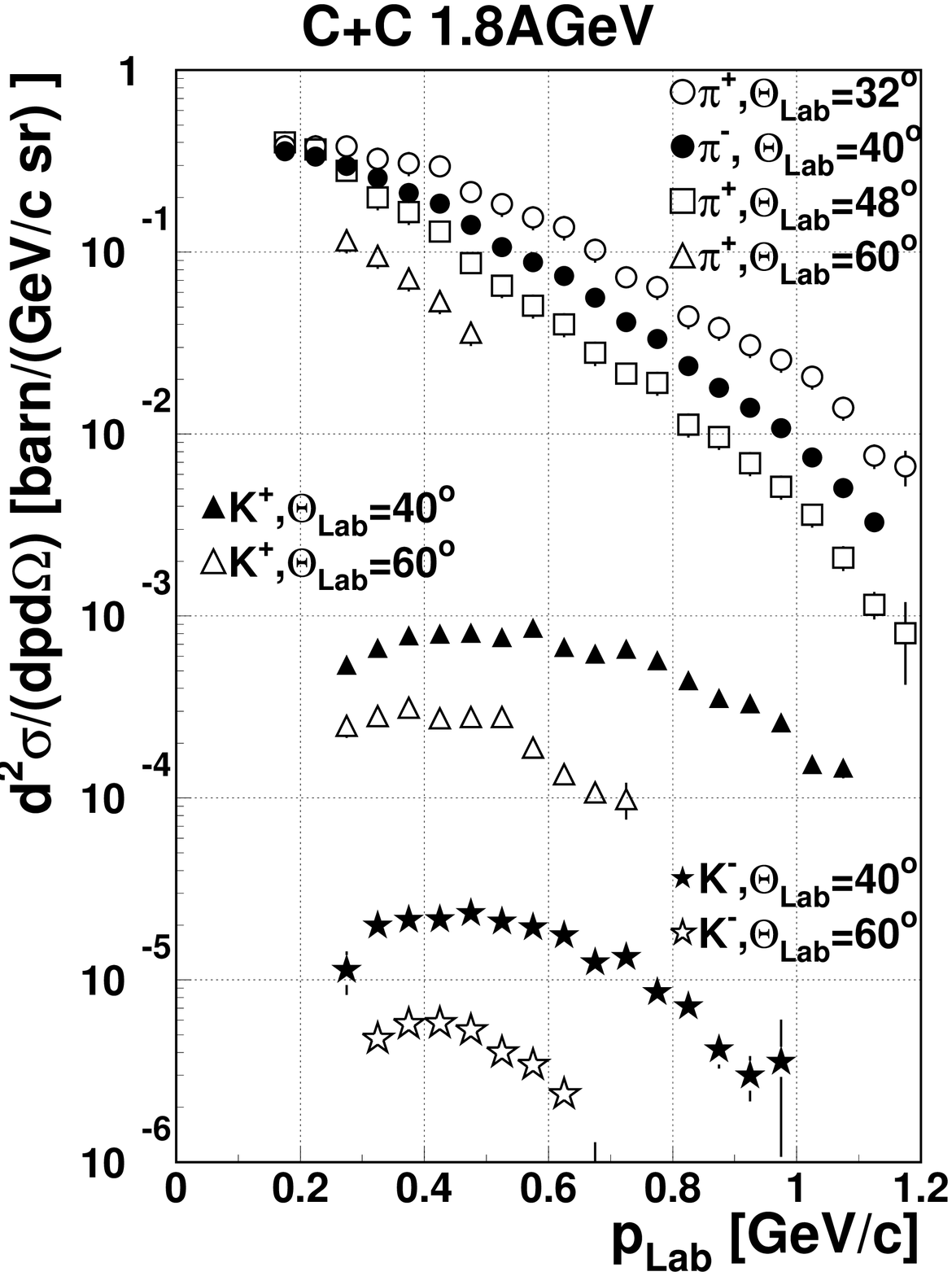,width=7.cm}}
\mbox{\epsfig{file=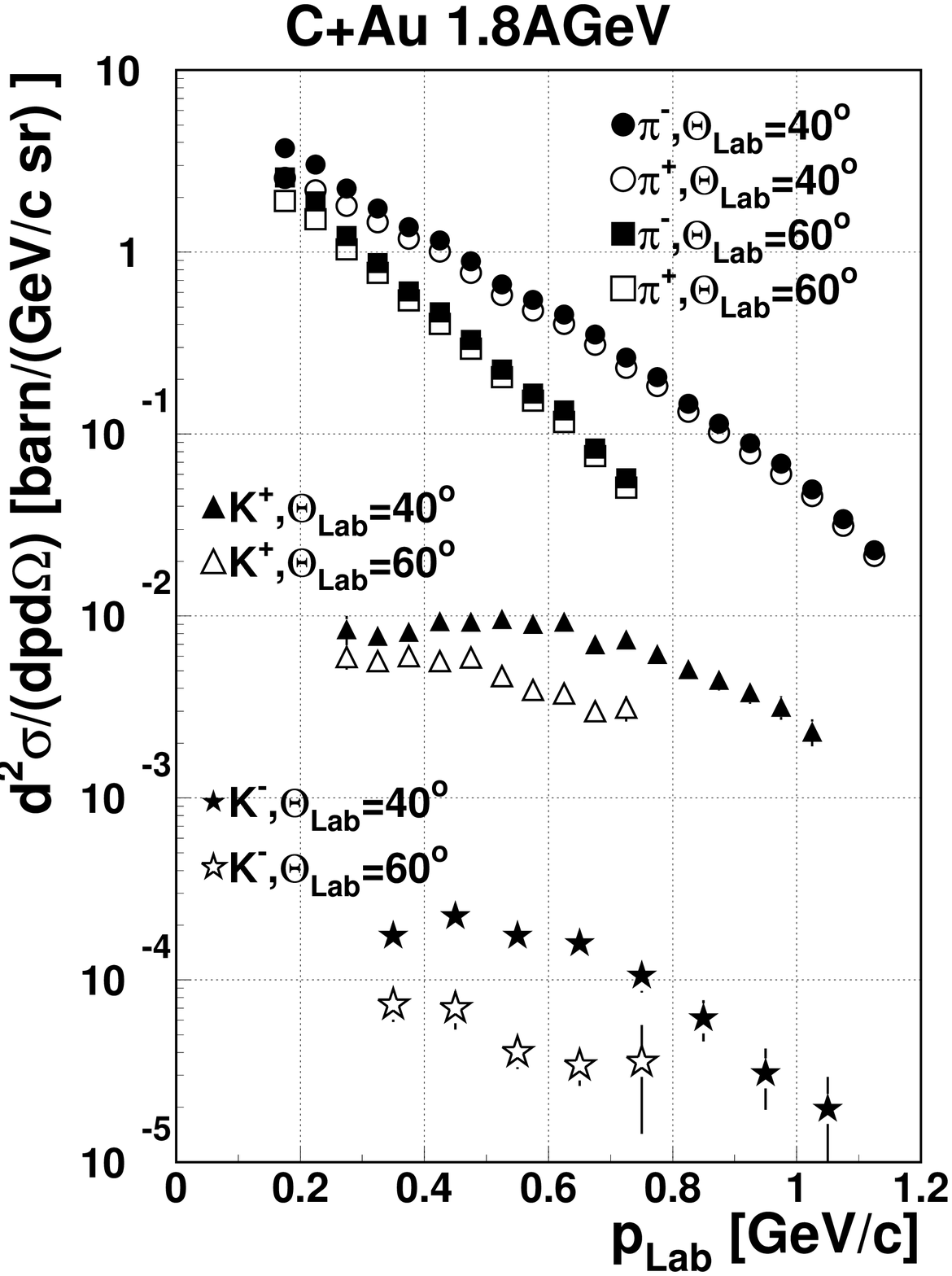,width=7.cm}}
}
\caption{Inclusive double-differential cross sections for the production of 
pions, K$^+$ mesons and
K$^-$ mesons measured in C+C (left) and C+Au (right) 
collisions at a beam energy of 1.8 AGeV
under different  laboratory angles (as indicated) as function of laboratory
momentum.
}
\label{data_cc18_lab}
\end{figure*}

Figures~\ref{data_cc10_lab} and \ref{data_cc18_lab}
present the double differential cross sections for the production of
pions, kaons and antikaons in C+C and C+Au 
collisions at a beam energy of 1 AGeV and 1.8 AGeV,   
as function of the laboratory momentum. 
Each particle species has been measured at least at two laboratory angles
in order to  obtain an estimate of the polar angle distribution.

In order to study the properties of the particle emitting source 
one has to determine the shape of the particle spectra and the angular
distributions in its particular Lorentz frame.  The transformation of the data 
into the center-of-mass of the participating nucleons 
is well defined only for symmetric systems. Therefore, we present in the
following only the data taken in C+C collisions.
The determination of the 
velocity of the particle emitting source in C+Au collisions will be 
discussed in the  next chapter.

\begin{figure*}[H]
\centerline{
\mbox{\epsfig{file=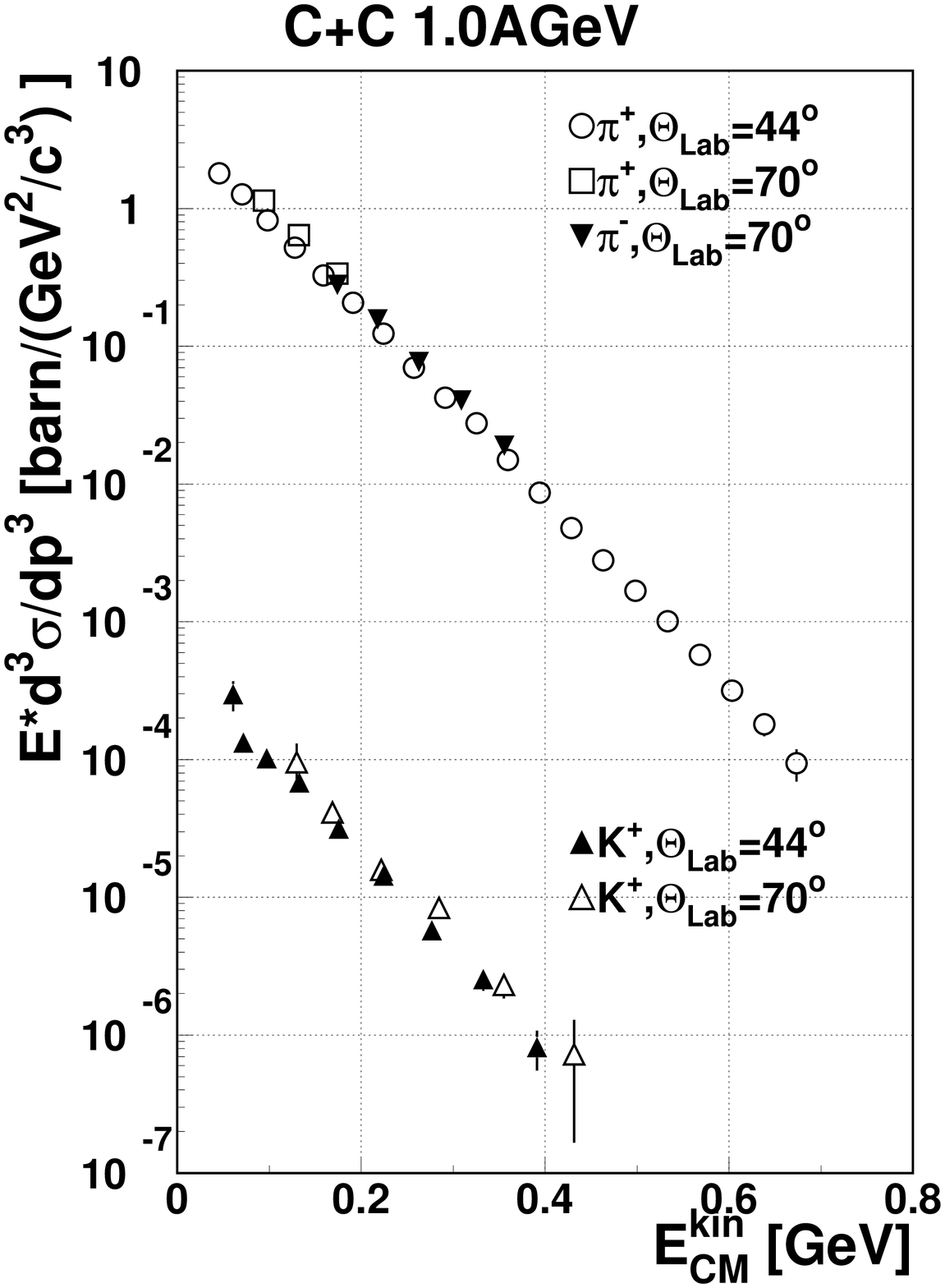,width=7.cm}}
\mbox{\epsfig{file=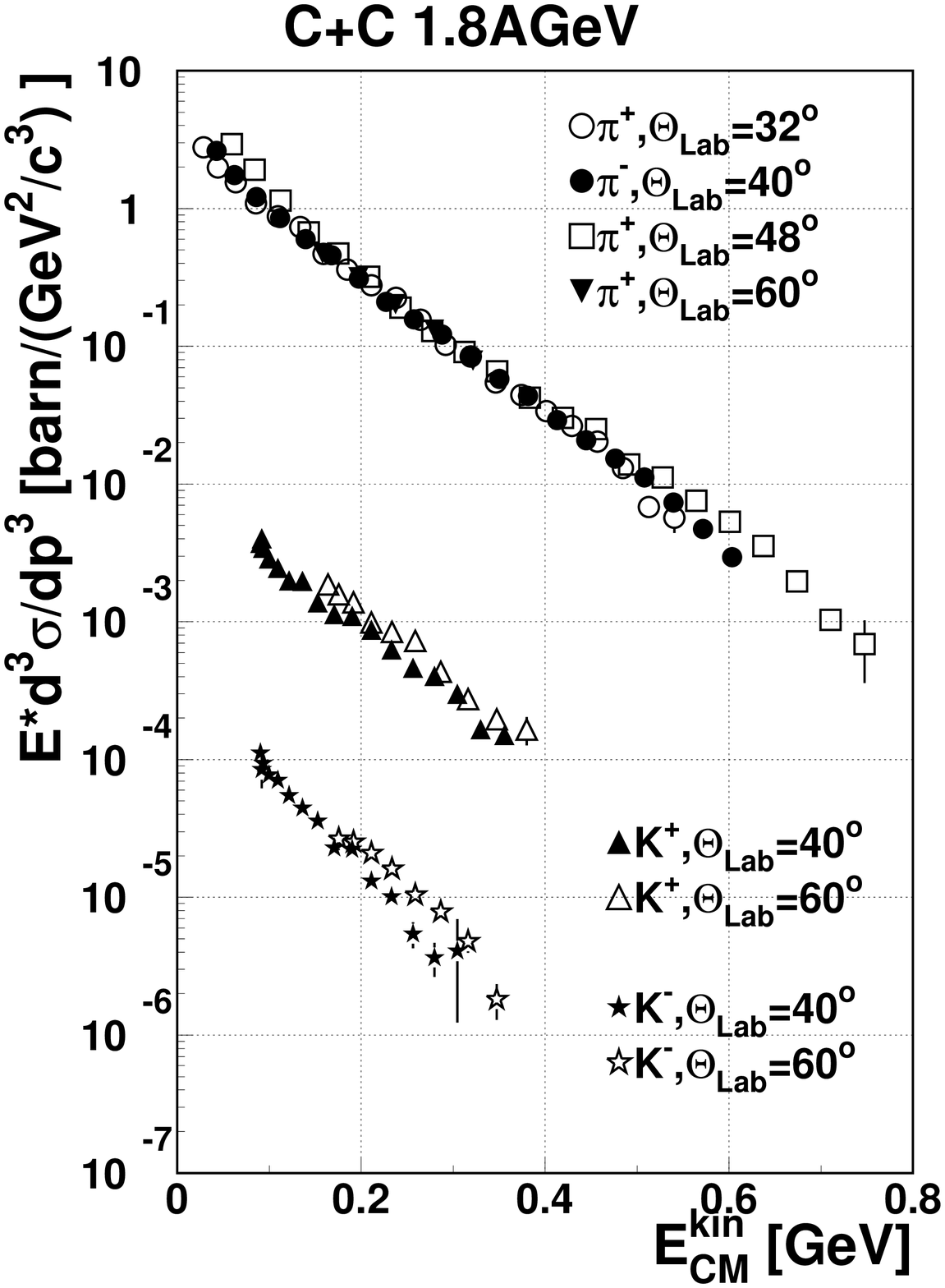,width=7.cm}}
}
\caption{Inclusive invariant cross sections for the production of pions,
K$^+$ mesons and
K$^-$ mesons measured in C+C collisions at a beam energy of 1.0 AGeV (left)
and 1.8 AGeV (right)
under different  laboratory angles (as indicated) as function of 
the c.m kinetic energy.
}
\label{data_cc_tcm}
\end{figure*}

Fig.~\ref{data_cc_tcm} shows the 
invariant production cross sections E d$^3\sigma$/dp$^3$ for 
pions, kaons and antikaons in C+C collisions
at 1.0 (left) and 1.8 AGeV (right) as function of their kinetic
energy in the c.m. system.  Boltzmann
distributions $d^3\sigma/dp^3 \propto exp(-E/T)$ have been fitted to the 
K$^+$ and K$^-$ spectra. 
In contrast, the pion spectra cannot be described well by a single 
Boltzmann distribution. Therefore, a superposition of two distributions 
has been fitted to the pion differential cross sections. The inverse 
slope parameters for the K mesons and for the high energy pions  
are listed in Table 2.

The K$^+$ (and K$^-$) spectra measured at different laboratory angles 
are slightly different when presented in the c.m. system.  
This indicates that the polar angle distribution is nonisotropic. 
A method to extract information on the angular distribution from our data
is discussed in the next  section.

\subsection{Polar angle distributions in C+C collisions}
In C+C collisions at 1.8 AGeV we have measured K$^+$ mesons at laboratory
angles of $\Theta_{lab} = 32^{\circ}, 40^{\circ}, 48^{\circ}$ and $60^{\circ}$. 
Figure~\ref{a2fit} shows the double differential cross sections as 
function of the  laboratory momentum. 
The dotted lines correspond to a simultaneous fit of a Boltzmann distribution 
$d^3\sigma/dp^3 \propto$ exp(-E/T) to the spectra assuming isotropic
emission. The fit results deviate systematically from the data: 
near midrapidity 
($\Theta_{lab} = 32^{\circ}$) 
the fit overestimates the data whereas at backward rapidities  
($\Theta_{lab} = 60^{\circ}$) the fit underestimates the data.  
In order to improve the agreement between fit and data we assume 
a forward-backward peaked polar angle distribution according to 
\begin{equation}
d^3\sigma/dp^3 \propto (1 + a_2 cos^2\Theta_{c.m.}) ~exp(-E/T)  
\label{polar}
\end{equation}
The solid lines represent  a simultaneous fit of this  function to the 
data. The resulting parameters are $a_2$ = 0.54$\pm$0.25 and T = 85$\pm$ 6 MeV
corresponding to a forward/backward peaked polar angle distribution.
The fraction of nonisotropically emitted kaons
can be determined by integrating equation (\ref{polar}) 
over full phase space:
\begin{equation}
\int dp^3 d^3\sigma/dp^3 \propto \int d\phi dcos\Theta (1 + a_2 
cos^2\Theta) \int  p^2 dp  ~exp(-E/T)
\label{polarint}
\end{equation}  
with $dp^3 = p^2 dp d\phi dcos\Theta$ and separation of the angular and
momentum dependence. The integration over the angles yields
$4\pi (1 + a_2/3)$. This means that the nonisotropic fraction of the total 
K$^+$ production cross section is only 
$(a_2/3)/(1 + a_2/3)$ = 0.15$\pm$0.07 for C+C collisions at 1.8 AGeV.

\subsection{Total production cross sections}

\begin{figure}[H]
\mbox{\epsfig{file=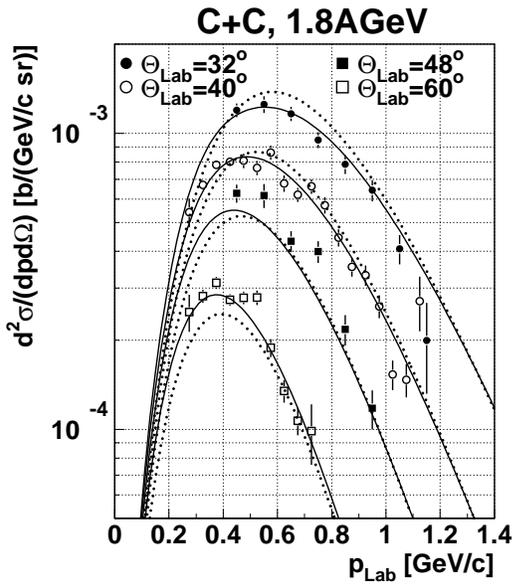,width=8.cm}}
\caption{Inclusive double-differential cross sections for the production
of K$^+$ mesons in C+C collisions at a beam energy of 1.8 AGeV  
under different  laboratory angles (as indicated) as function of 
the laboratory momentum.
The lines represent a simultaneous fit of a Boltzmann distribution
to the data assuming an isotropic (dotted lines) and a nonisotropic
(solid lines) polar angle distribution (see text).  
}
\label{a2fit}
\end{figure}

When integrating the double-differential cross-sections for K$^+$
production in C+C collisions at 1.8 AGeV (as presented in fig.~\ref{a2fit})
over the full phase space assuming an isotropic emission in 
the c.m. system, one obtains for the total cross section the values 
$\sigma_{K^+}$ = 2.4$\pm$0.3 mb ($\Theta_{lab}$=32$^{\circ}$),
$\sigma_{K^+}$ = 2.9$\pm$0.3 mb ($\Theta_{lab}$=40$^{\circ}$),
$\sigma_{K^+}$ = 3.3$\pm$0.3 mb ($\Theta_{lab}$=48$^{\circ}$) and
$\sigma_{K^+}$ = 3.6$\pm$0.4 mb ($\Theta_{lab}$=60$^{\circ}$). 
The different results demonstrate the influence of the polar angle distribution.

Total meson production cross sections can be extracted from our data 
by integrating the fit function in Eq.~\ref{polar} over full 
phase space. Using this procedure we obtain for    
the total K$^+$ production cross section in C+C collisions at 1.8 AGeV 
a value of $\sigma_{K^+}$ = 3.0$\pm$0.3 mb. 

A second method for the determination of the production cross section 
is based on  (i) the assumption of a parabolic
polar angle distribution (according to Eq.~\ref{polar})
and on (ii) the observation that for a certain laboratory 
angle an isotropic particle source  and a nonisotropic source 
yield the same differential cross section.  
This matching angle is about $\Theta_{lab}$=42$^{\circ}$ 
for K$^+$ mesons from C+C production at 1.8 AGeV  
(see fig.~\ref{a2fit}). Therefore, the total cross sections
calculated for  different laboratory angles 
(assuming isotropic emission, see above) can be interpolated 
to the angle of $\Theta_{lab}$=42$^{\circ}$. The result 
is $\sigma_{K^+}$ = 3.0$\pm$0.3 mb, exactly the same value as for the 
simultaneous fit of a nonisotropic angular distribution. 
The second method has the advantage that the result of the interpolation 
is largely independent of the value of $a_2$.  Therefore, the error of 
$a_2$ which might be very large when only two laboratory angles have been 
measured, influences the result only weakly.  
The matching angle depends
only on the beam energy and on the assumption of a parabolic 
polar angle distribution.

Table 2  contains the total  production cross sections
for pions, kaons and antikaons measured in C+C and C+Au collisions 
at 1.0 and 1.8 AGeV. The calculation of the cross sections for
the asymmetric C+Au system requires information on the velocity of 
the particle emitting source (see next section).

\subsection{The kaon emitting source in C+Au collisions}

In order to study the properties of the fireball the particle observables
have  to be analysed in the rest frame of the source. 
For a symmetric collision system, this frame coincides with the 
center-of-mass frame of the nuclei (i.e. the  nucleon-nucleon c.m. frame).
For the asymmetric C+Au system the situation is much less clear. 
In a first step, one can roughly estimate the number of the 
participating nucleons (and their origin) for the C+Au system
by a geometrical model assuming 
straight line geometry \cite{huefner}  
\begin{equation}      
<A_{part}> = \frac{A_P\pi R^2_T + A_T\pi R^2_P}{\pi(R_P + R_T)^2}
\label{apart}
\end{equation}
with $A_P$ and $A_T$ the mass number and $R_P$ and $R_T$ the radius of 
projectile and target nucleus, respectively. For symmetric 
collisions systems the average number of participants is $<A_{part}>$ = A/2,
i.e. $<A_{part}>$ = 6 for $^{12}$C+$^{12}$C collisions.
For $^{12}$C+$^{197}$Au collisions 
the average number of participants 
(for inclusive reactions) is  $<A_{part}>$ = 22 with 
6.6 participants from the projectile and 15.4 from the target nucleus
according to equation~\ref{apart}.

However, with respect to particle production the number of participating 
nucleons from the target nucleus might differ from the geometrical   
estimate. 
Therefore, we have extracted the velocity of the particle emitting 
source from a transport model calculation. E.L. Bratkowskaya and W. Cassing
have calculated K$^+$ production in C+Au collisions at 1.8 AGeV
with their RBUU code \cite{cass_brat} and made us the events available. 
The calculation uses a parameterization of 
elementary cross sections which is based on data including those 
recently measured at  COSY \cite{balewski} and takes into account 
rescattering and in-medium effects.
The predictions of the model calculation agree well with the 
differential cross sections for K$^+$ and K$^-$ production in 
C+C and C+Au collisions (see section 4).
We have analysed the momentum distribution of the kaons and have determined
the reference frame in which the average longitudinal momentum of the kaons 
is zero. This frame moves with the velocity of a 
fireball which contains twice as many nucleons 
from the target than from the projectile nucleus 
(e.g. 6 projectile nucleons and 12 target nucleons).    
The fireball velocity is $\beta$ = 0.44 for the C+Au system at 1 AGeV 
(as compared to $\beta$ = 0.59 for C+C at 1 AGeV) and 
 $\beta$ = 0.56 for the C+Au system at 1.8 AGeV
(as compared to $\beta$ = 0.70 for C+C). 

\begin{figure*}[H]
\centerline{
\mbox{\epsfig{file=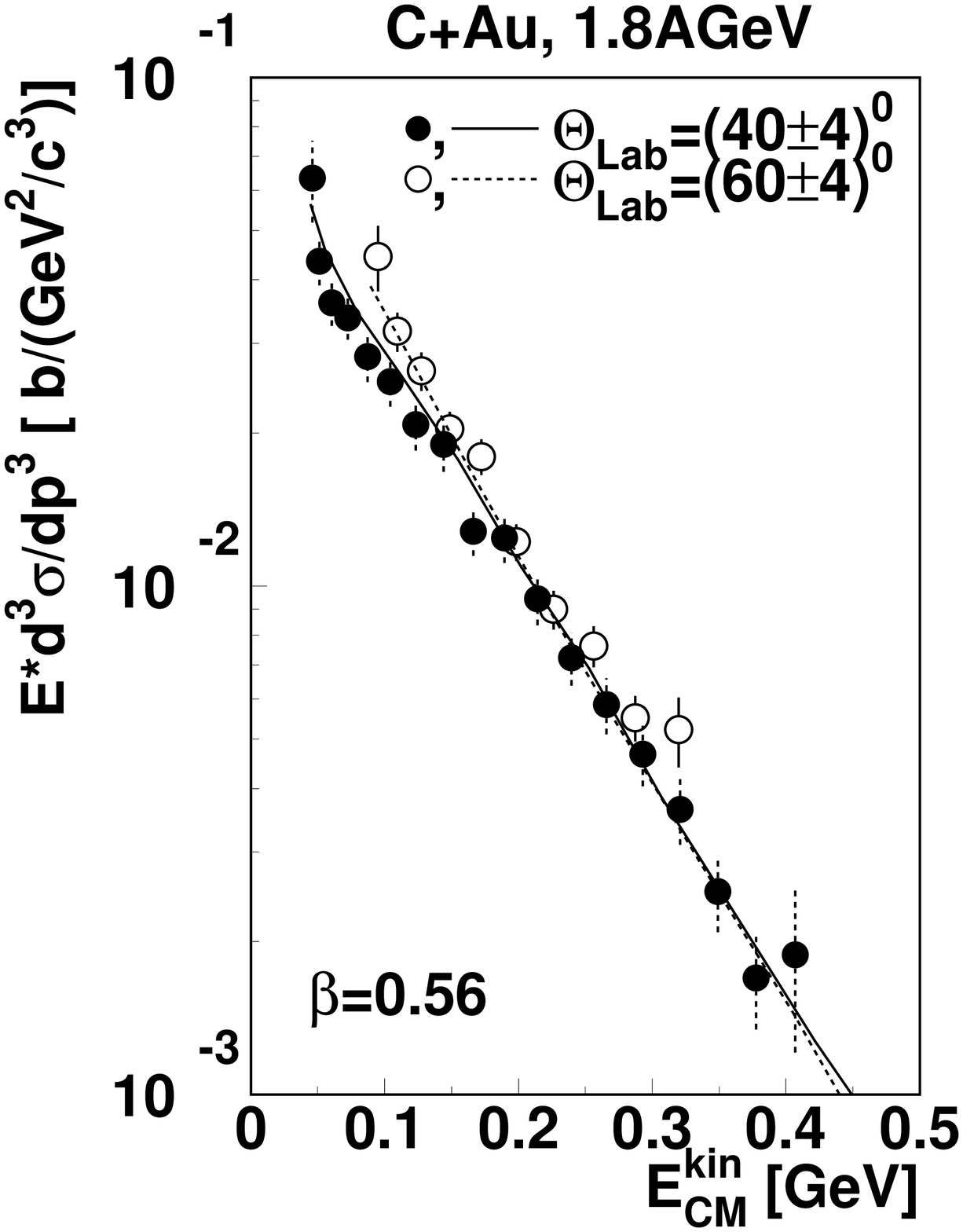,width=7.cm}}
\mbox{\epsfig{file=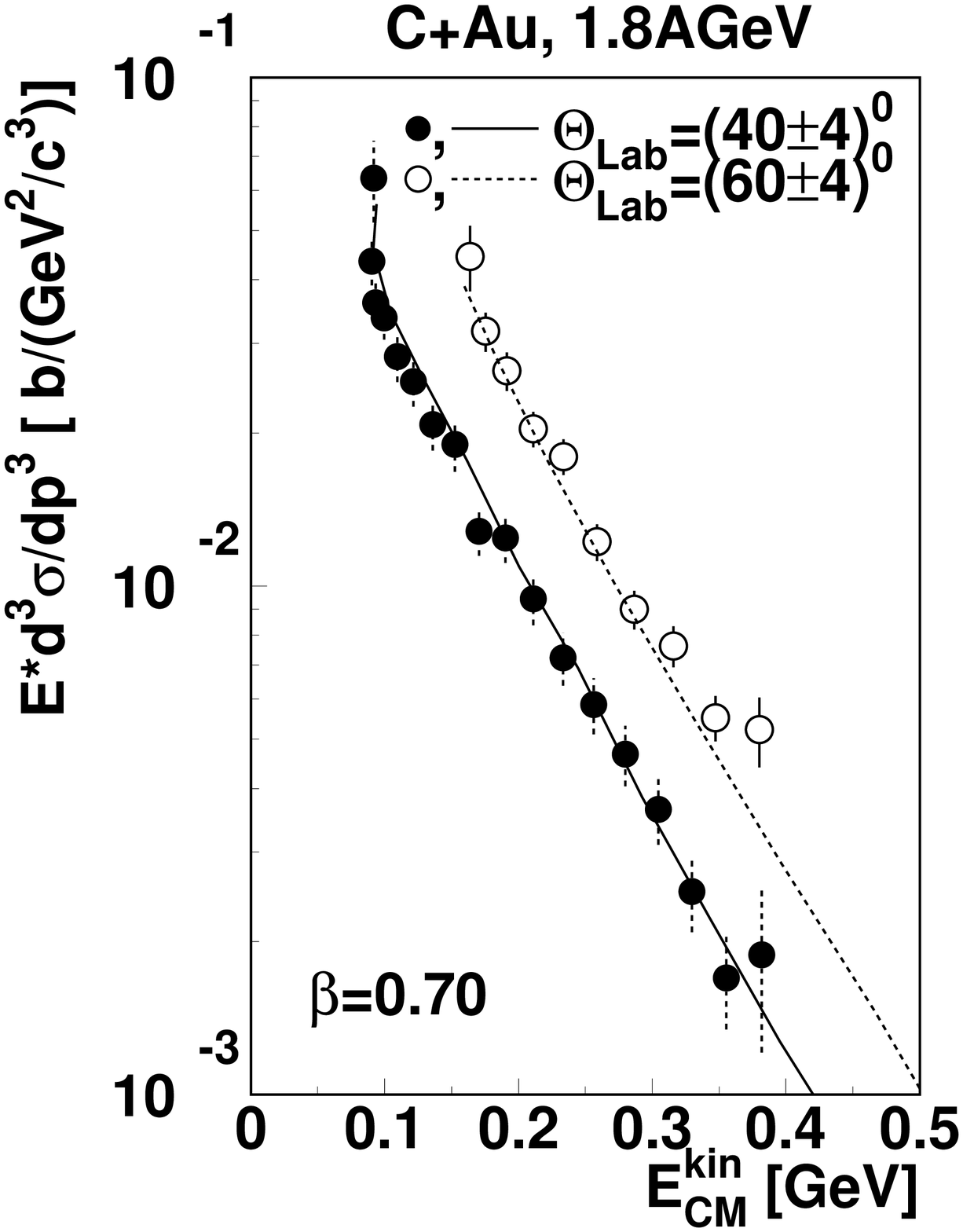,width=7.cm}}
}
\caption{ Inclusive invariant cross sections for the production 
of K$^+$ mesons measured in C+Au collisions at a beam energy of 1.8 AGeV 
as function of the kinetic energy in different  source frames.
The source velocity is assumed to be either $\beta$ = 0.70 (corresponding 
to the nucleon-nucleon system, right figure) or 
$\beta$ = 0.56 (corresponding to a source which contains twice as much 
nucleons from the target nucleus than from the projectile, left figure)
The data (full/open symbols) and the RBUU predictions (solid/dashed lines)
are taken at $\Theta_{lab}$ = $40^{\circ}$/60$^{\circ}$. 
}
\label{source}
\end{figure*}

Within these source frames both kaons and antikaons are emitted 
nearly isotropically.  
This is demonstrated for K$^+$ mesons in the left part of   
fig.~\ref{source} both for  RBUU predictions and data. 
Figure \ref{source} (left) 
presents the invariant cross sections for  K$^+$ production
in C+Au collisions at 1.8 AGeV 
transformed into the frame of the source which contains twice as much 
nucleons from the target nucleus than from the projectile.
The K$^+$ spectra taken at
$\Theta_{lab}$ = $40^{\circ}$ and 60$^{\circ}$ 
nearly coincide in this representation. In contrast, when 
transforming measured and calculated spectra into the nucleon-nucleon frame
(e.g. a source containing 6 nucleons from the projectile and 6 from the target)
the polar angle distribution appears very much nonisotropic (see
right part of fig.~\ref{source}).

\begin{figure*}[H]
\centerline{
\mbox{\epsfig{file=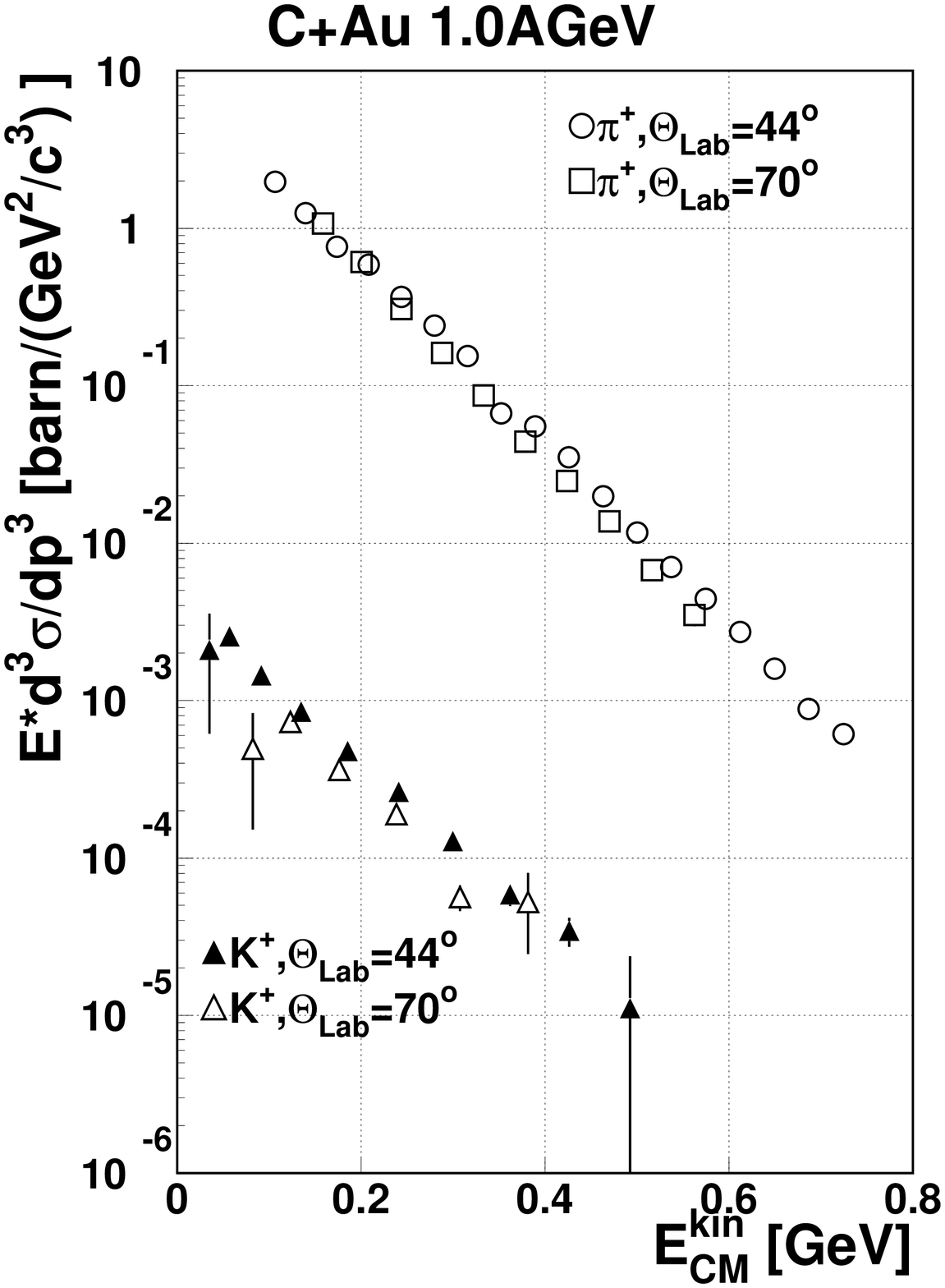,width=7.cm}}
\mbox{\epsfig{file=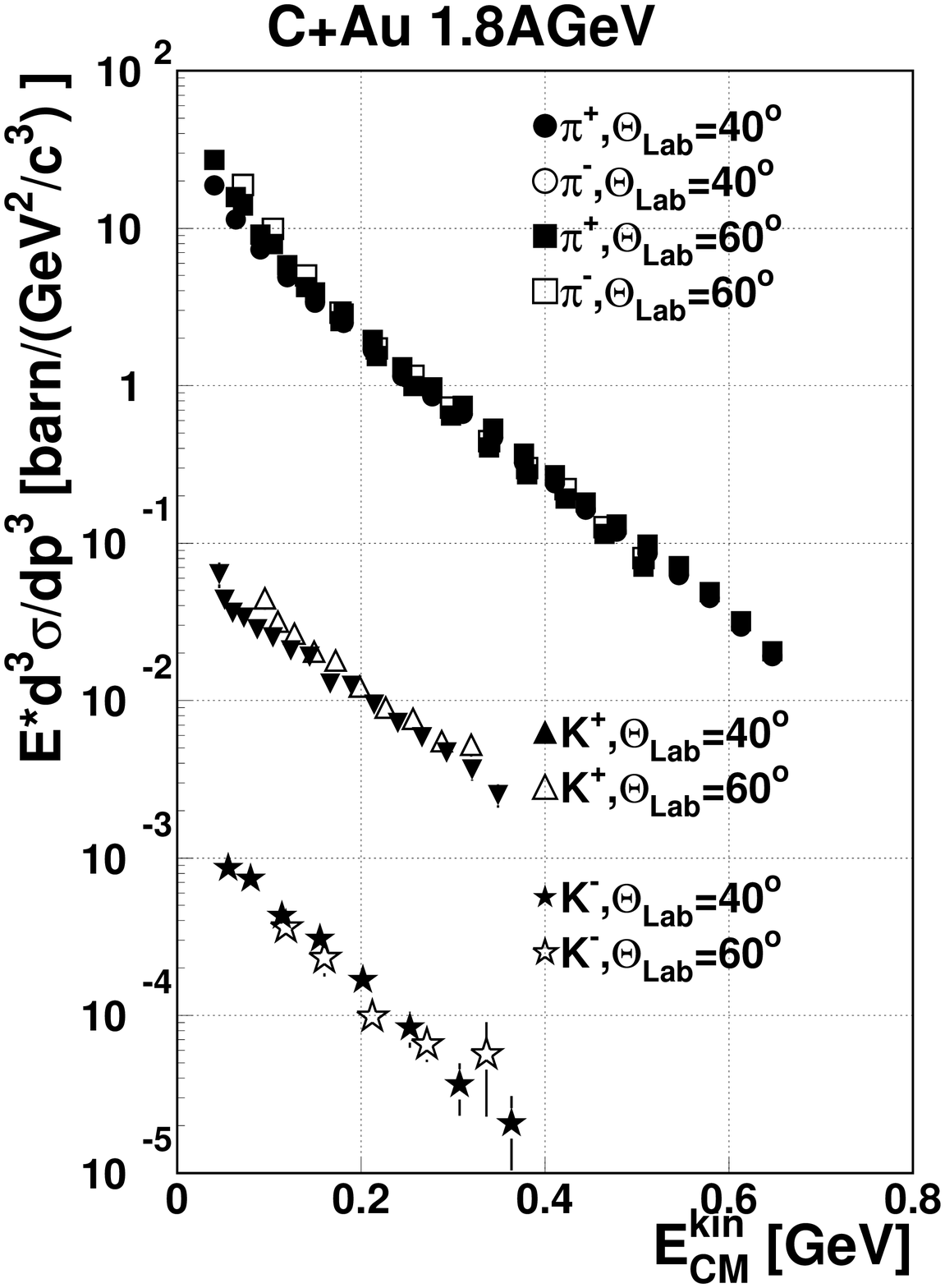,width=7.cm}}
}
\caption{Inclusive invariant cross-sections for the production of pions,
K$^+$ mesons and
K$^-$ mesons measured in C+Au collisions at a beam energy of 1.0 AGeV (left)
and 1.8 AGeV (right) as function of the kinetic energy in the source frame. 
The source moves with the velocity of the center-of-mass of a system 
containing  one  projectile nucleon and two target nucleons.  
}
\label{data_cau_tcm}
\end{figure*}

In order to determine spectral distributions and multiplicities 
of pions, kaons and antikaons from C+Au collisions 
we adopt  source velocities of  $\beta$ = 0.45 and 0.56 
for beam energies of 1.0 and 1.8 AGeV, respectively, as
given by the RBUU transport calculation for K$^+$ and K$^-$ mesons. 
Fig.~\ref{data_cau_tcm} presents the invariant meson production cross sections 
in C+Au collisions at 1.0 AGeV (left) and 1.8 AGeV (right) as function
of the kinetic energy in the source frame.  
Boltzmann distributions have been fitted to the data (not shown in 
fig.~\ref{data_cau_tcm}). 
The resulting inverse slope parameters  are given in Table 2.

\subsection{Particle multiplicities}
Inclusive cross sections for the production of pions, kaons and antikaons
in C+Au collisions have been determined from Fig.~\ref{data_cau_tcm}.
The Boltzmann distributions are used to  extrapolate over the non-measured 
regions of phase space taking into account the angular distributions 
as measured. The cross sections are listed in Table 2.    
The multiplicity of produced particles per collision can be calculated by 
M = $\sigma/\sigma_R$ with $\sigma$ the  production cross section. 
The geometrical reaction cross section for a collision of two nuclei
with masses $A_P$ and $A_T$ is defined as
\begin{equation}
\sigma_R = \pi(r_0 A_P^{1/3} + r_0 A_T^{1/3})^2 
\label{react}
\end{equation}
with $r_0$ = 1.2 fm. This definition
yields $\sigma_R$= 0.95 barn for the $^{12}$C+$^{12}$C system and 
$\sigma_R$= 3.0 barn for $^{12}$C+$^{197}$Au collisions.     
Table 2 includes the particle multiplicities per participating nucleon 
M/$<A_{part}>$. The average number of participating nucleons
$<A_{part}>$ is calculated for inclusive reactions according to 
equation~\ref{apart}.

It is interesting to note that the pion multiplicity per
participating nucleon M($\pi^+)/<A_{part}>$ is a factor of more than three
smaller in C+Au than in C+C collisions at 1 AGeV (the effect of 
isospin might  change this factor slightly, in C+Au collisions at 1.8 AGeV  
the $\pi^+/\pi^-$ ratio is 0.84$\pm$0.18). The TAPS Collaboration 
measured neutral pions in C+C and C+Au collisions at 0.8 AGeV (at midrapidity
in the NN system) and found 
M($\pi^0)/<A_{part}>$ reduced by a factor of about 2 when using the heavy
target \cite{averbeck}. 
At a beam energy of 1.8 AGeV, however, 
the pion multiplicities per participating 
nucleon in C+C and C+Au collisions differ only by a factor of about 0.7 
(averaged value for $\pi^+$ and $\pi^-$). 

The strong reduction of the pion multiplicity per participant in C+Au 
collisions at 1 AGeV as compared to the C+C system indicates 
that pions are reabsorbed in the large gold nucleus. 
When increasing  the projectile energy, 
the projectile nucleons will collide subsequently 
with nucleons in the Au target 
at energies  well above the pion production threshold. This effect 
increases the number of produced pions and thus compensates partly for
absorption. Such a scenario explains - 
at least  qualitatively - the observation that the pion multiplicity
per participating nucleon rises faster with beam energy  in C+Au than in 
C+C collisions and hence is almost similar 
for both targets  at a beam energy of 1.8 AGeV  (see Table 2).

In contrast to the pions, 
the values of the K$^+$ multiplicities per nucleon M(K$^+$)/$<A_{part}>$
are very similar for C+C and C+Au collisions at 1 AGeV and at 1.8 AGeV.
From symmetric collision systems at beam energies
below or near the K$^+$ production threshold it is known that  
M(K$^+$) increases more than linearly  with $<A_{part}>$ \cite{barth,laue}. 
The similar values  of 
M(K$^+$)/$<A_{part}>$ for C+C and C+Au indicate that the energy 
per participating nucleon available for particle production in C+Au 
collisions is smaller than in the C+C system. 

In Ni+Ni collisions at 1.8 AGeV it was found that the  
multiplicities of K$^-$ and K$^+$ mesons scale similarly with $<A_{part}>$
\cite{barth}. Therefore, one would expect also an enhancement 
of M(K$^-$)/$<A_{part}>$ for C+Au as compared to C+C collisions at 1.8 AGeV.   
However, we observe that the K$^-$ multiplicity per participating nucleon
is reduced by a factor of 0.55 in the C+Au system. This result indicates
a significant loss of K$^-$ mesons in C+Au collisions 
which overcompensates for the expected enhancement. The loss  
may be caused by reabsorption of antikaons in the heavy target nucleus
via the strangeness exchange  reaction K$^-$N$\to\Lambda\pi$. 
The inelastic K$^-$p cross section 
(at K$^-$ momenta of 300 MeV/c) is about 40 mb \cite{dover}.
This corresponds to a mean free path of about $\lambda\approx$1.5 fm  
for K$^-$ mesons
in nuclear matter at saturation density ($\rho_0$ = 0.17 fm$^{-3}$).

Our data indicate  that the absorption of  antikaons is an  
important process in nucleus-nucleus collisions. This effect has to be
understood quantitatively if one tries to extract information 
on the in-medium properties of antikaons from heavy-ion data. 
In the small C+C collision system, however, 
K$^-$ absorption should be of minor importance and in-medium effects 
might be observable. 

\begin{figure}[H]
\mbox{\epsfig{file=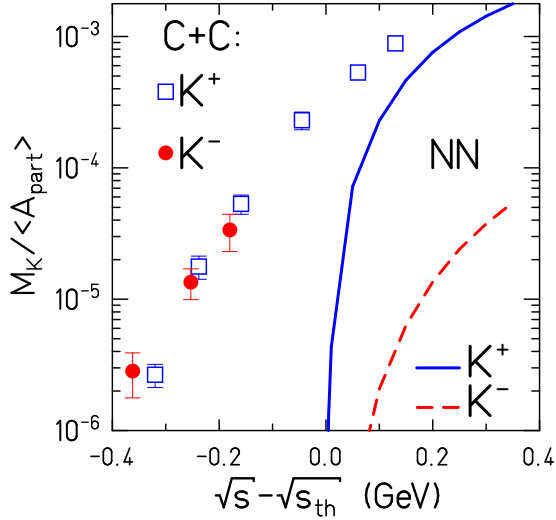,width=10.cm}}
\caption{Kaon and antikaon multiplicity per participating nucleon
as a function of the Q-value  for C+C
collisions (open squares: K$^+$, full dots: K$^-$) and
nucleon-nucleon  collisions.
The error bars include systematic effects.
The lines (NN) correspond to parameterizations of the  isospin averaged
cross sections for K meson production measured in proton-proton collisions
(full line: K$^+$, dashed line: K$^-$)
\protect\cite{si_ca_ko,brat_cass,sibirtsev}.
}
\label{CC_KP_KM_EXCI}
\end{figure}

A way to visualize the effect of the nuclear
medium on the production of kaons and antikaons is proposed in 
fig.~\ref{CC_KP_KM_EXCI} (taken from \cite{laue}).    
The figure  presents the K$^+$ and K$^-$ multiplicities per participating
nucleon for C+C and nucleon-nucleon collisions 
as function of the  energy above threshold
($\sqrt s$-$\sqrt s_{th}$) in the nucleon-nucleon (NN) system.
The values of the excess energy $\sqrt s$-$\sqrt s_{th}$
are calculated according to
$s$ = (E$_{pro}$+2m$_N$)$^2$ - p$_{pro}^2$ with E$_{pro}$ and p$_{pro}$
the projectile kinetic  energy and momentum per nucleon and m$_N$
the nucleon mass. The threshold energy
is $\sqrt s_{th}$=m$_K$+m$_{\Lambda}$+m$_N$ = 2.55 GeV 
for K$^+$ production and
$\sqrt s_{th}$=2m$_K$+2m$_N$ = 2.86 GeV for K$^-$K$^+$ pair production.

The solid and the dashed line in 
fig.~\ref{CC_KP_KM_EXCI}  represent the
parameterizations of the isospin averaged  
cross sections for K$^+$ and K$^-$ production
in nucleon-nucleon collisions \cite{si_ca_ko,brat_cass,sibirtsev}.
These calculations reproduce the available
experimental cross sections including  data measured recently 
close to threshold at
COSY and SATURNE  \cite{balewski,balestra}.
The multiplicities as shown in fig.~\ref{CC_KP_KM_EXCI} 
 are calculated  from
the elementary cross sections using $\sigma_R$ = 45 mb
and A$_{part}$=2 for nucleon-nucleon collisions.

The data in fig.~\ref{CC_KP_KM_EXCI} 
demonstrate that the excitation functions
for K$^+$ and K$^-$ production in C+C collisions are quite similar
when correcting the energy axis  for the threshold energies.
In nucleon-nucleon (NN)
collisions, however, the K$^+$ yield exceeds the K$^-$ yield by 1-2 orders of
magnitude for beam energies close to threshold.
The comparison of the K$^+$ and K$^-$ excitation functions for C+C and
nucleon-nucleon reactions near threshold clearly indicates that
in the nuclear medium the K$^-$ multiplicity is much more 
enhanced  than the K$^+$ multiplicity. Possible reasons for the enhancement
of K$^-$ production in nucleus-nucleus collisions 
will be discussed in the next section.

\section{Medium effects in kaon and antikaon production}

In nucleus-nucleus collisions at bombarding energies near the  
production threshold,  strange mesons are created  predominantly  
in secondary processes like $\pi$N$\to$K$^+$Y, $\Delta$N$\to$K$^+$YN,
$\pi$N$\to$K$^-$K$^+$N,  $\Delta$N$\to$K$^-$K$^+$NN and $\pi$Y$\to$K$^-$N
with Y=$\Lambda,\Sigma$.
These processes represent mechanisms of energy accumulation via multiple 
collisions. Moreover, the effective particle production threshold
is lowered by the Fermi motion of the nucleons. Therefore, kaons and 
antikaons are observed in C+C collisions below the nucleon-nucleon threshold
of $\sqrt s$-$\sqrt s_{th}$=0 (see fig.~\ref{CC_KP_KM_EXCI}). 
According to relativistic transport models the pion and $\Delta$ induced
sequential processes dominate the K$^+$ production at bombarding energies
near the kinematical threshold \cite{cassing,fuchs}.
In the case of K$^-$ mesons, however, the most important channel is
expected to be the strangeness exchange reaction $\pi$Y$\to$K$^-$N. 
However, the calculations which take into account these 
secondary K$^-$ production channels (including K$^-$ absorption)
underpredict the K$^-$ yield measured in Ni+Ni collisions at 1.8 AGeV 
\cite{barth} by a factor of about 6 when neglecting
in-medium mass modifications of K$^-$ mesons \cite{cassing}.

\subsection{K$^+$ and K$^-$ production at equivalent beam energies}

One would expect that density dependent in-medium effects are more pronounced
in Ni+Ni than in C+C collisions. As illustrated in fig.~\ref{CC_KP_KM_EXCI}
medium effects can be quantified for example by the measured ratio
K$^-$(1.8 AGeV)/K$^+$(1.0 AGeV). These two energies are ''equivalent'' in
the sense that they correspond to very similar (negative) values for the 
excess energy of $\sqrt s$-$\sqrt s_{th}$ $\approx$ -0.23 GeV. 
However, this ratio is 1$\pm$0.4 for Ni+Ni \cite{barth} and 
0.76$\pm$0.25 for C+C \cite{laue}, 
i.e. there is no difference within the error bars.
On the other hand, the reabsorption of antikaons  should be enhanced in 
larger systems. This effect may cancel the enhancement due to the 
in-medium effect with the consequence that the K$^-$/K$^+$ ratio at 
equivalent energies is independent of the size of the collision system.

\begin{figure*}[H]
\centerline{
\mbox{\epsfig{file=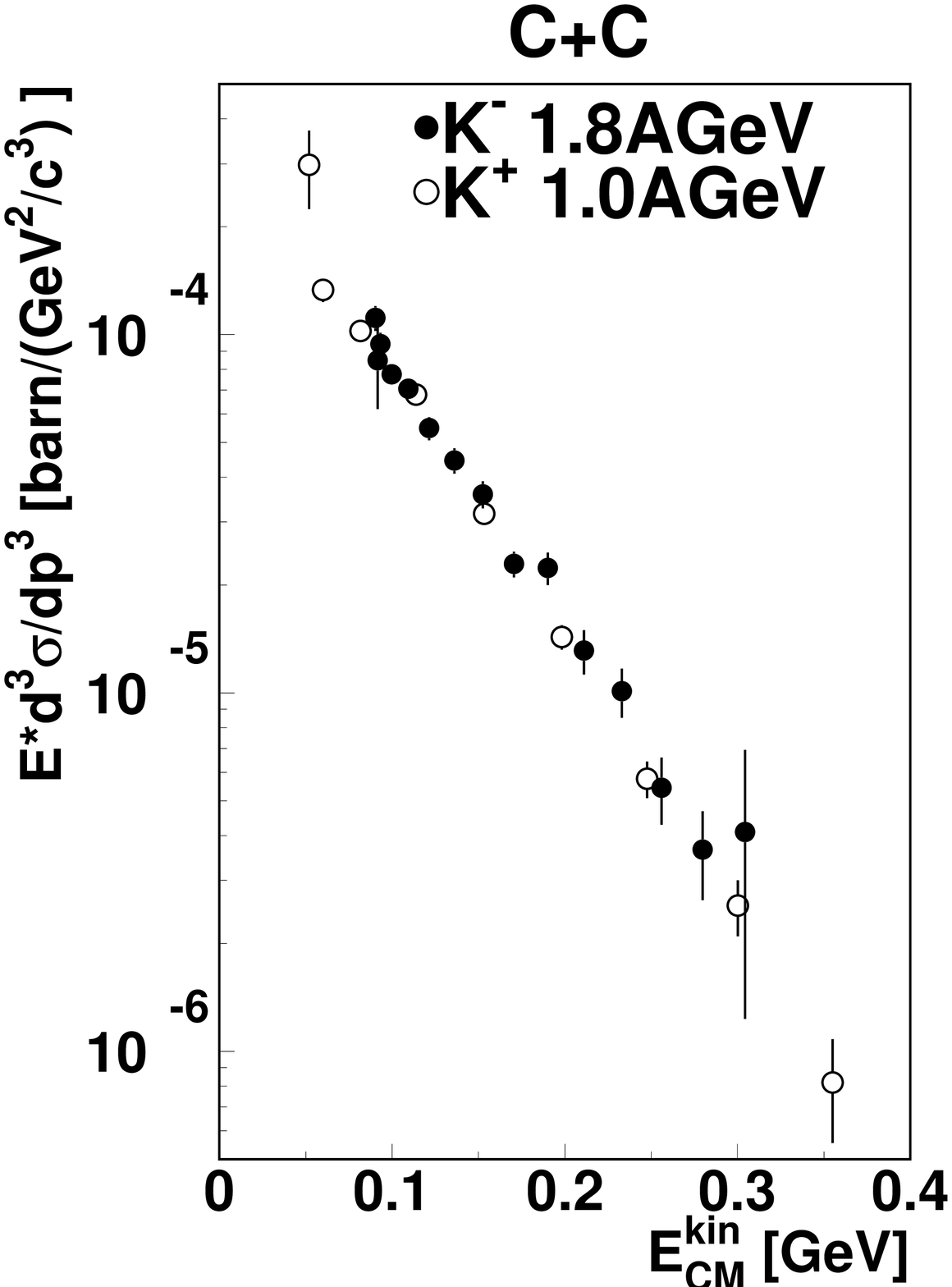,width=7.cm}}
\mbox{\epsfig{file=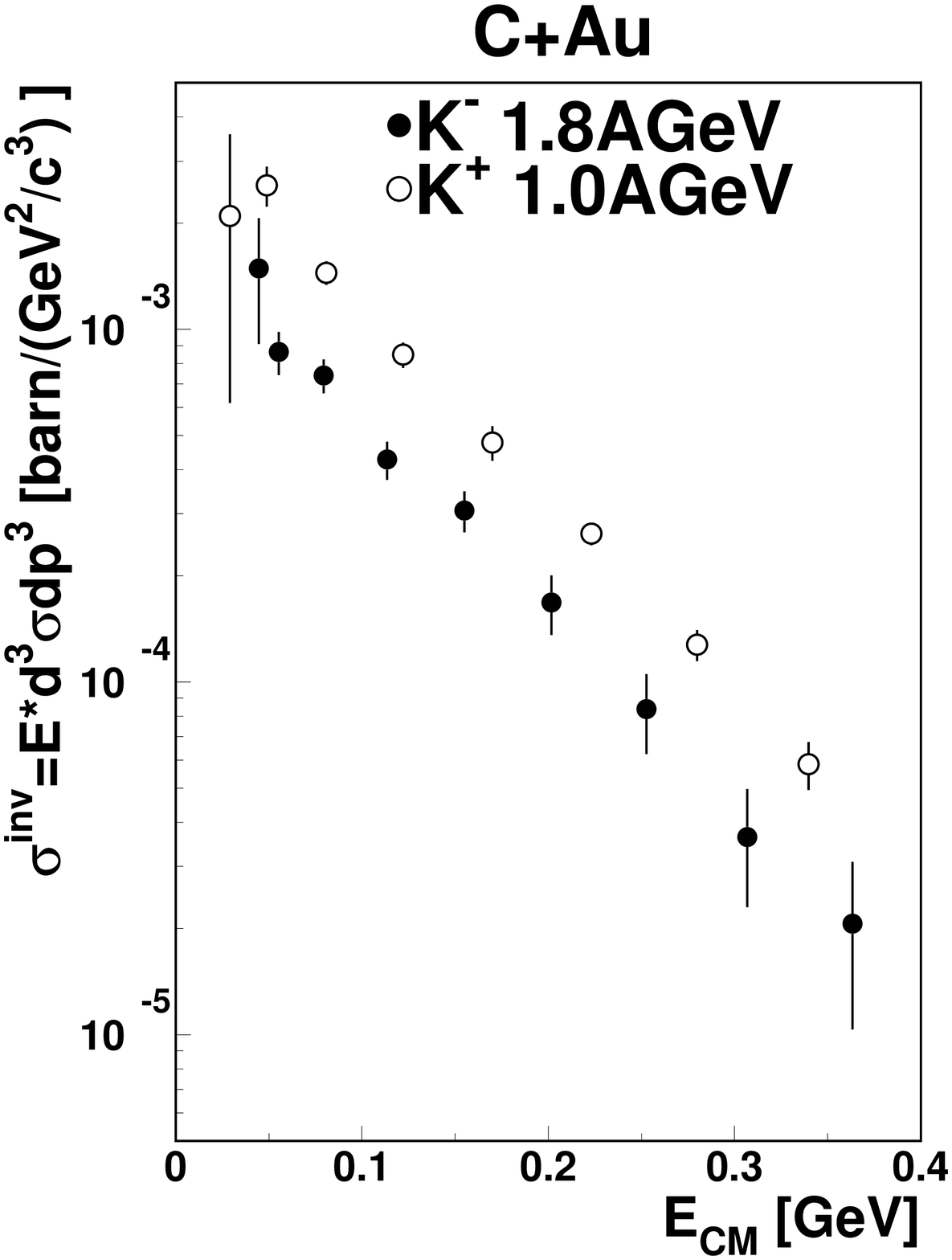,width=7.cm}}
}
\caption{Comparison of invariant cross-sections for  K$^-$ and K$^+$ 
production at equivalent energies (see text) 
in C+C (left) and C+Au collisions (right).
}
\label{equival}
\end{figure*}

Fig.~\ref{equival} presents the invariant production cross sections
of K$^+$ and K$^-$ mesons  as function of c.m. kinetic energy 
for C+C (left) and C+Au (right) collisions at ''equivalent'' beam energies.
In C+C collisions, the K$^+$ yield measured at 1.0 AGeV agrees to the
K$^-$ yield taken at 1.8 AGeV similar as for Ni+Ni.
In C+Au collisions, however,  the K$^-$ data are reduced 
with respect to the  K$^+$ data by more than  factor of 2.
This observation may be explained by reabsorption of antikaons 
in the target spectator nucleus. 

\subsection{Comparison to transport model calculations}
In the following we will compare the kaon and antikaon data 
to predictions of RBUU
transport calculations \cite{cass_brat}.  
The model takes into account the in-medium modification 
of the effective masses of strange mesons according to 
m$^*_K$ = m$^0_K$ ( 1 - $a \rho/\rho_0$) with m$^0_K$ the kaon mass in 
the vacuum, $\rho$ the baryonic density inside the reaction zone,  
$\rho_0$ the saturation density and $a$ = -0.06 for kaons and 
$a$ = 0.24 for antikaons. The values of $a$ correspond to a slight increase
of the kaon mass with increasing baryonic density (12\% at $\rho = 2\rho_0$)
and a significant decrease of the effective antikaon mass 
(48\% at $\rho = 2\rho_0$). 
The assumption on the linear density dependence of the kaon and antikaon masses 
reproduces a trend  found by various calculations \cite{waas,schaffner}.
It should be mentioned that the effect of dynamical spectral functions  
on the in-medium properties of K mesons \cite{lutz}
is ignored in this simple parameterization.

The variation of the  effective masses  
change the in-medium production thresholds and thus the kaon and antikaon
yields. At beam energies below and near the thresholds, 
the kaon and antikaon excitation
function rises steeply with increasing bombarding energy and thus 
the  kaon and antikaon yields depend very sensitively upon the in-medium
thresholds.  
     
\begin{figure*}[H]
\centerline{
\mbox{\epsfig{file=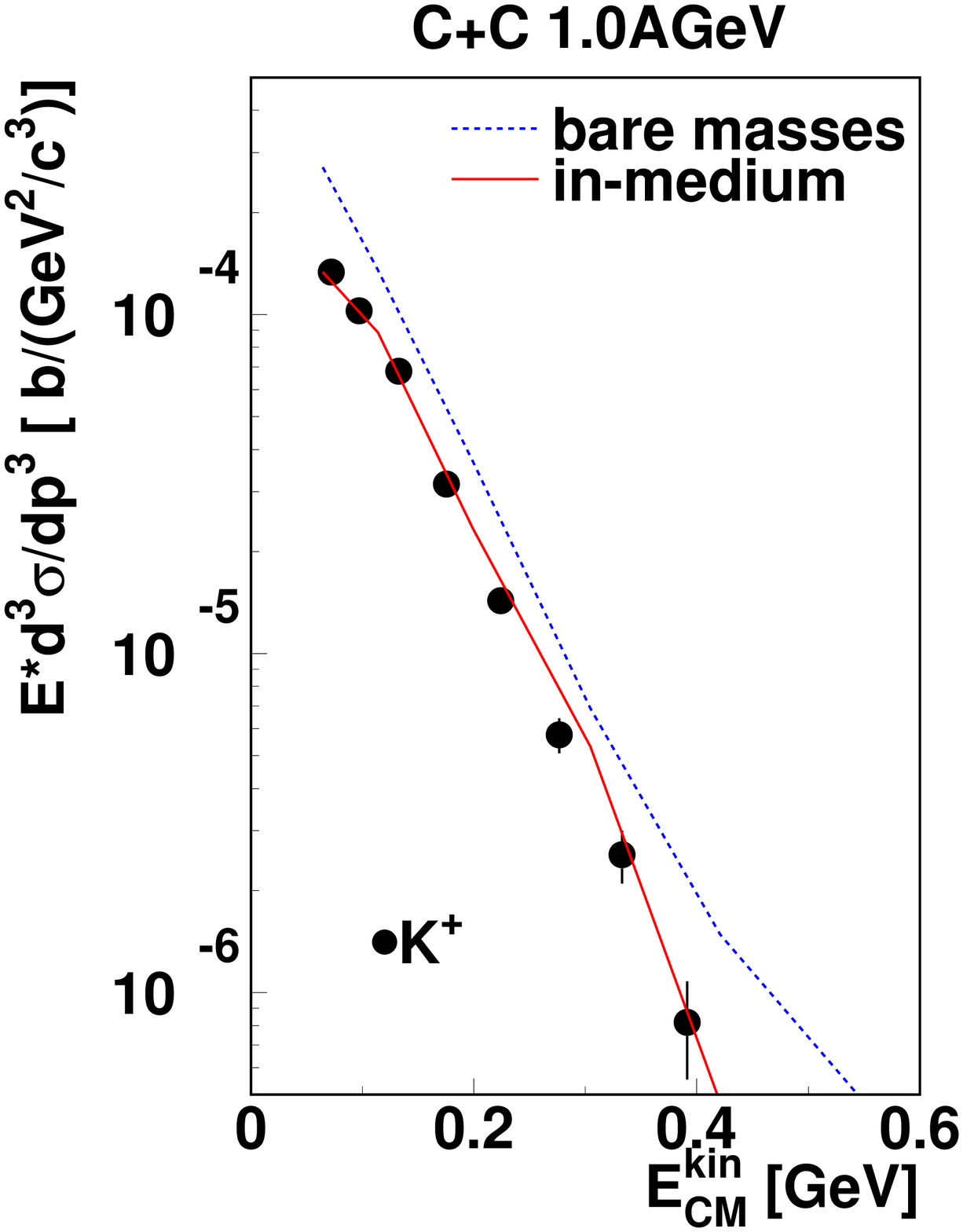,width=7.cm}}
\mbox{\epsfig{file=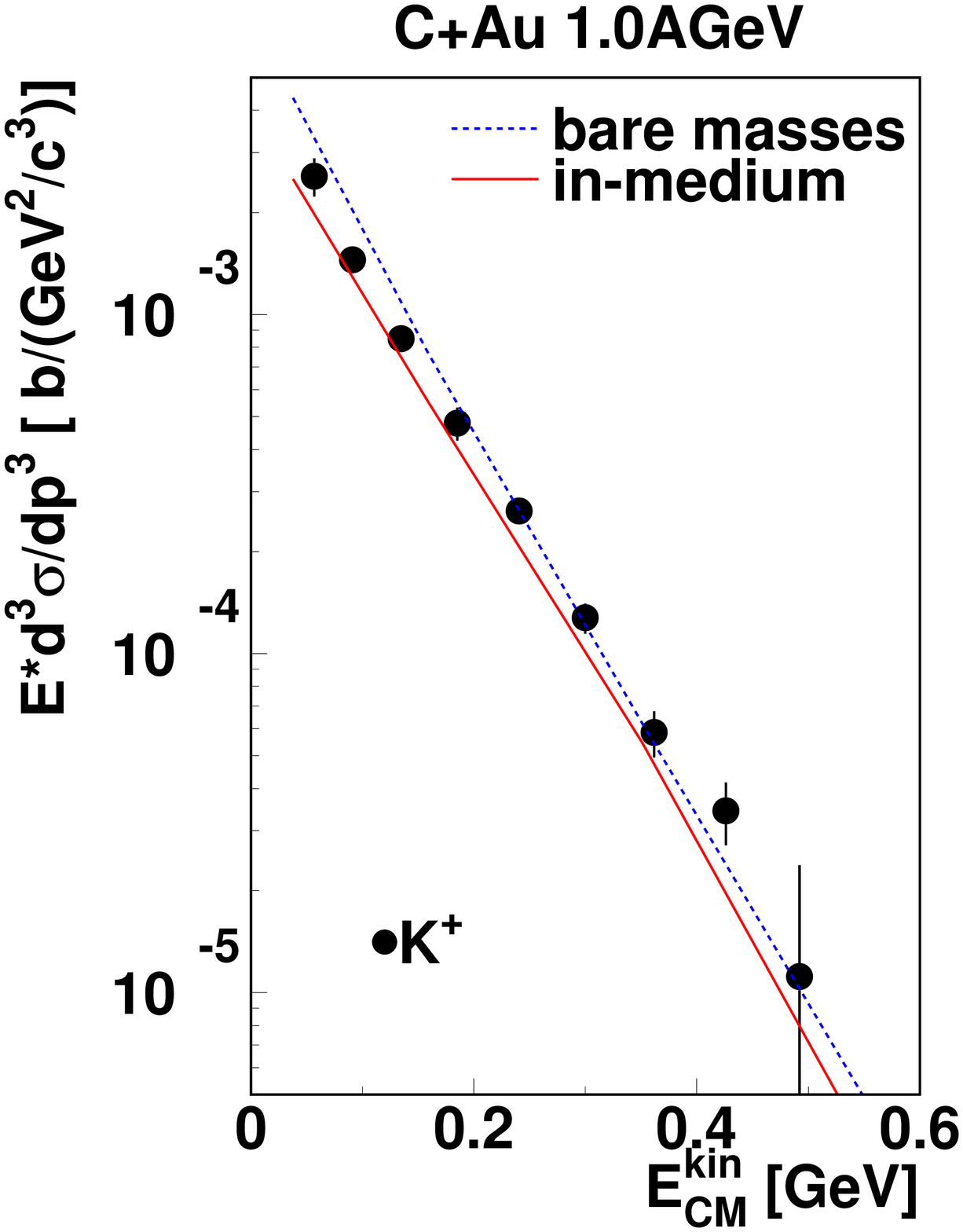,width=7.cm}}
}
\caption{ Invariant cross sections for K$^+$ production in C+C (left)
and C+Au collisions (right) at 1.0 AGeV. The data are compared 
to results of a RBUU transport calculation with (solid lines) and without 
(dashed lines) in-medium modifications of the K$^+$ mesons 
\protect\cite{brat_priv}.
}
\label{kp10buu}
\end{figure*}

\begin{figure*}[H]
\centerline{
\mbox{\epsfig{file=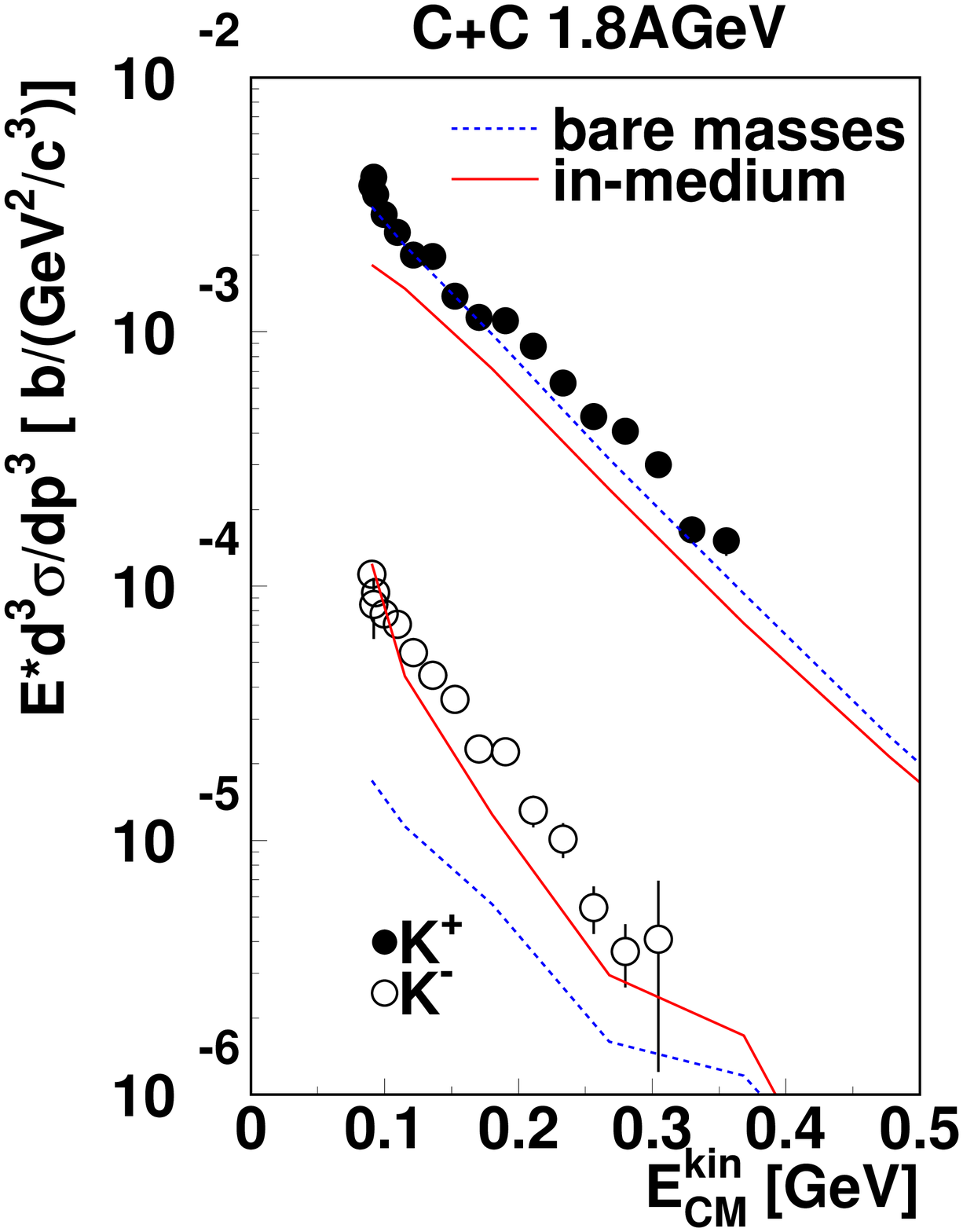,width=7.cm}}
\mbox{\epsfig{file=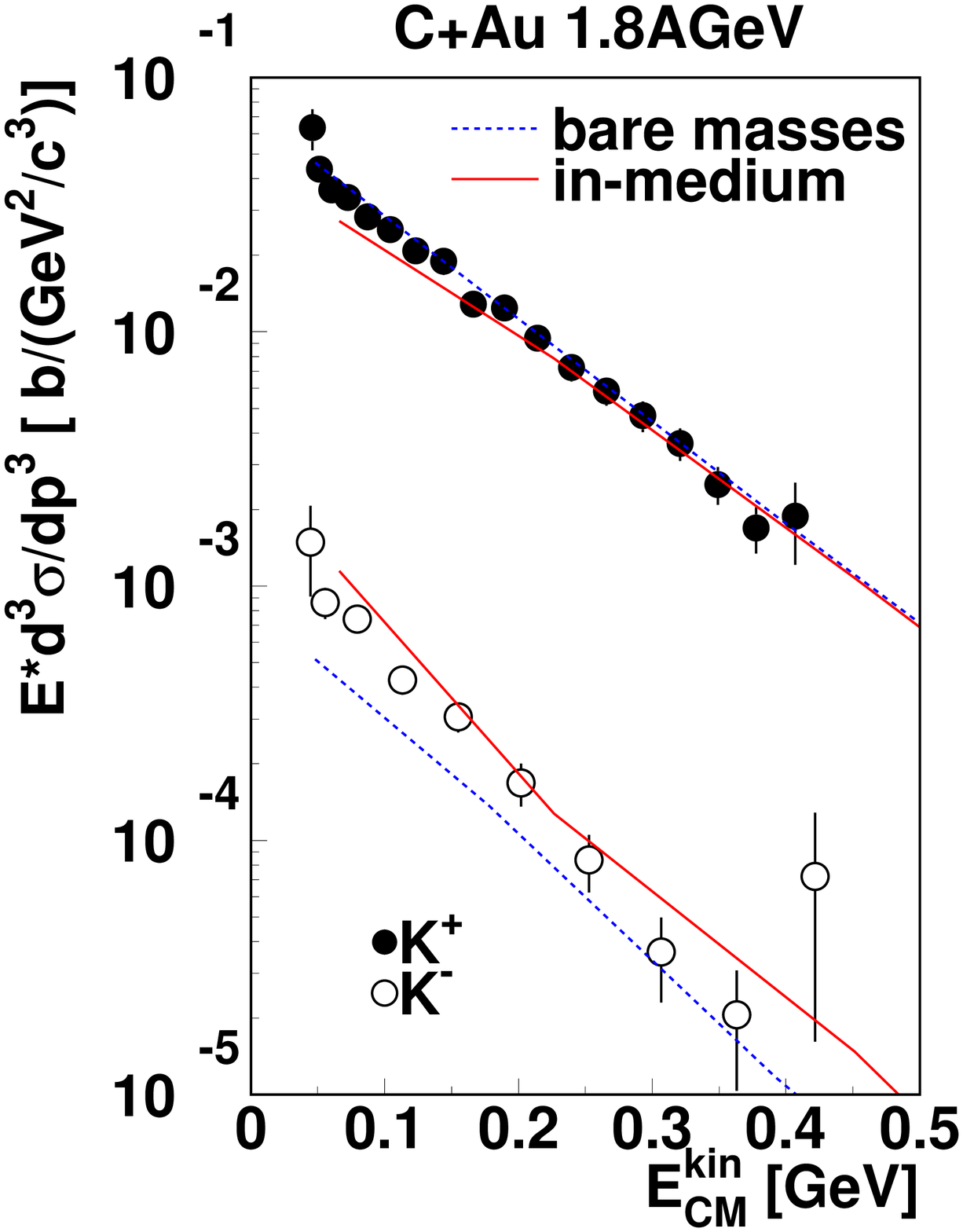,width=7.cm}}
}
\caption{Invariant cross sections for K$^+$ and K$^-$ production in C+C (left)
and C+Au collisions (right) at 1.8 AGeV. The data are compared 
to results of a RBUU transport calculation with (solid lines) and without 
(dashed lines) in-medium modifications of the K mesons 
\protect\cite{brat_priv}.
}
\label{kpkmbuu}
\end{figure*}

In fig.~\ref{kp10buu} and fig.~\ref{kpkmbuu} the measured kaon and 
antikaon  invariant production cross sections
are compared to predictions of RBUU transport calculations \cite{brat_priv}.
The solid lines correspond to the result of the transport code if 
in-medium effects are taken into account. 
The dashed lines represent the calculation assuming ''bare'' masses.
For C+C collisions at 1.0 AGeV (fig.~\ref{kp10buu}, left part)
the kaon data clearly favor the ''in-medium'' calculation 
in contrast to the kaon data measured in the same system at 1.8 AGeV 
(fig.~\ref{kpkmbuu}, left part).
In C+Au collisions both at 1.0 AGeV (fig.~\ref{kp10buu}, right part)
and at 1.8 AGeV (fig.~\ref{kpkmbuu}, right part)
the data cannot distinguish the two calculations.   
Hence, no clear signature of in-medium effects on K$^+$ production 
emerges from the comparison of RBUU calculations and experiment. However, 
transport calculations performed recently with two QMD codes \cite{ai_fu}
are able to reproduce K$^+$ production cross sections measured in
C+C collisions at bombarding energies between 0.8 and 2.0 AGeV only 
when taking account a repulsive KN potential.

On the other hand,  the RBUU model calculation predicts an enhanced 
K$^-$ yield when in-medium effects are taken into account
(see fig.~\ref{kpkmbuu}).
For C+C collisions at 1.8 AGeV, 
the K$^-$ differential cross section exceeds the
''bare mass'' calculation by about a factor of 5 whereas the 
enhancement is less pronounced in C+Au collisions at 1.8 AGeV.    
Hence, the  measured antikaon yield is much better (although not perfectly)
described by the calculations with in-medium effects included.

\begin{figure*}[H]
\centerline{
\mbox{\epsfig{file=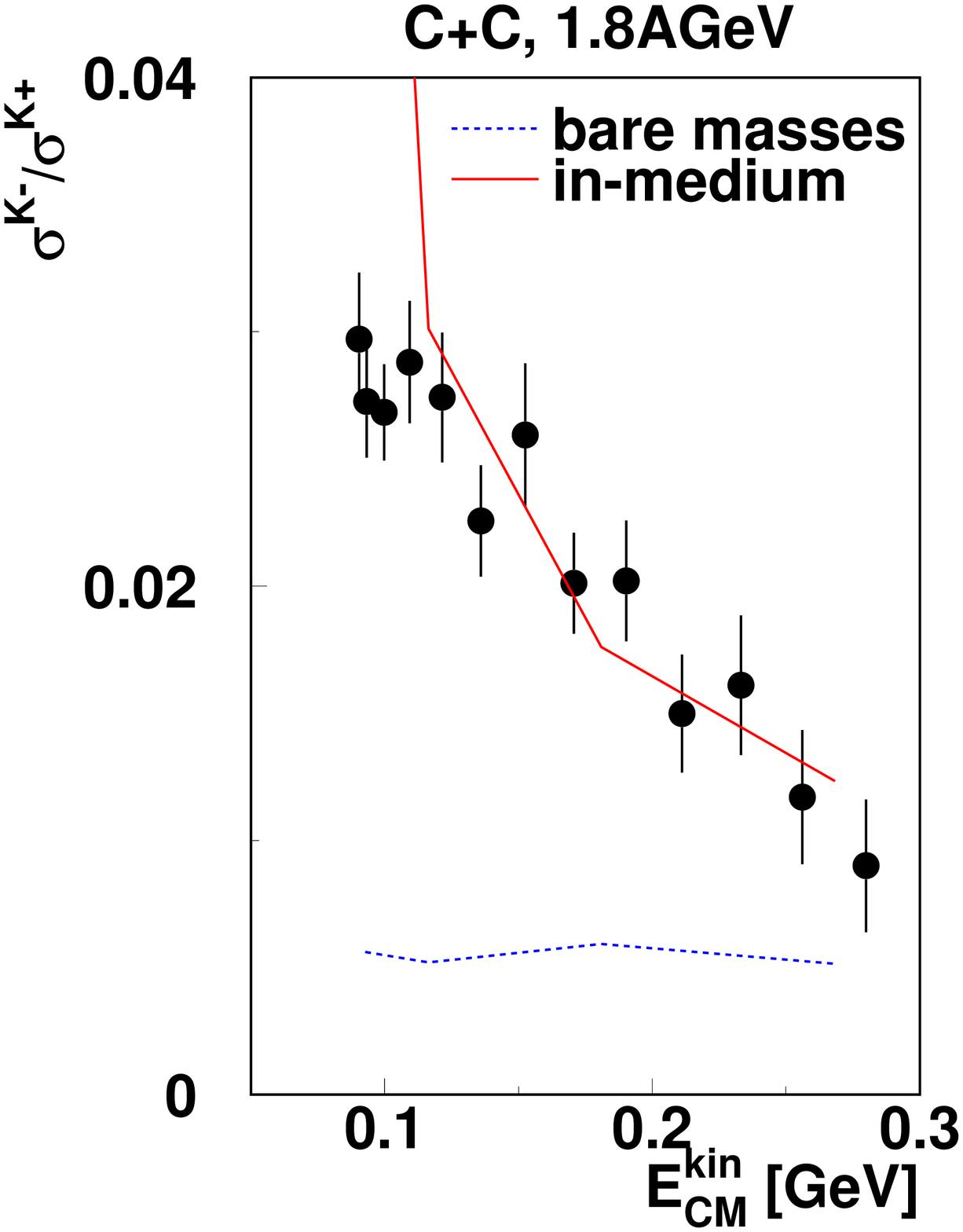,width=7.cm}}
\mbox{\epsfig{file=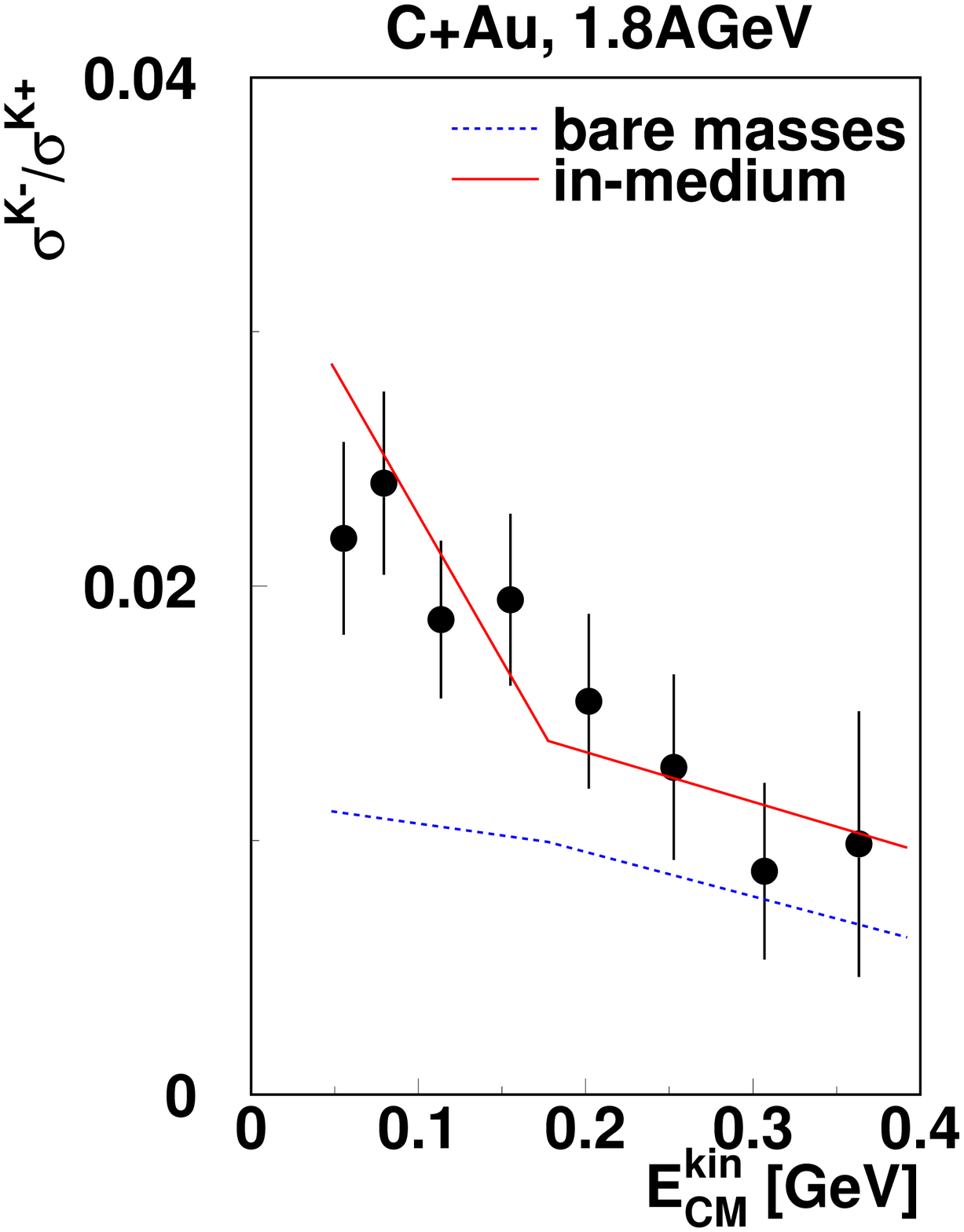 ,width=7.cm}}
}
\caption{K$^-$/K$^+$ ratio for C+C (left) and C+Au collisions (right)
at 1.8 AGeV. The data are compared 
to results of a RBUU transport calculation without (solid lines) and with 
(dashed lines) in-medium modifications of the K mesons 
\protect\cite{brat_priv}. The data shown in the left figure 
have been published in \protect\cite{laue}.
}
\label{karat}
\end{figure*}

As mentioned before, the measured cross sections are affected by 
an uncertainty of 15\% due to systematic errors. Moreover, 
the results of the model calculations depend in absolute yield 
on the treatment of momentum dependent interactions, the role of 
pions and baryonic resonances, the nuclear equation of state etc.. 
In order to reduce these uncertainties we present in fig.~\ref{karat} 
the K$^-$/K$^+$ ratios in C+C collisions (left) and C+Au collisions (right)
at a beam energy of 1.8 AGeV as function of the c.m. kinetic energy.
The calculations predict two important features of the data:    
the relative yields and spectral slopes  of kaons and  antikaons.
According to the RBUU calculation, the steeper spectral slope of the 
antikaons with respect to the kaons is caused by the in-medium 
kaon-nucleon potentials which are repulsive for kaons but attractive 
for antikaons. The number of calculated antikaons is much smaller
than the number of measured antikaons. Therefore, we have
grouped the calculated K mesons into 3-4 bins only. This presentation 
causes the discontinuities of the model results.

\section{Conclusions}
We have presented production cross sections of charged pions, 
kaons and antikaons measured in C+C and C+Au collisions 
at beam energies of 1.0 and 1.8 AGeV at different polar emission angles. 
For the asymmetric $^{12}$C+$^{197}$Au 
system we have determined the velocity of the K meson emitting
source by transport calculations. It turns out that the source moves with 
the velocity of the center-of-mass  of one nucleon from the projectile 
and two
nucleons from the target. Within this frame both the calculated and the 
measured K mesons exhibit essentially an isotropic emision pattern.   

The kaon and antikaon spectra are described by Boltzmann distributions
whereas the pion spectra exhibit an additional enhancement at low energies.
The spectral slopes are similar for high-energy pions and kaons 
but are steeper for antikaons.
The inverse slope parameters increase both with beam energy and target size.

At a beam energy of 1 AGeV, the pion multiplicity per participating nucleon  
M($\pi^+$)/$<A_{part}>$ is about a factor of three smaller 
in C+Au than in C+C collisions whereas at 1.8 AGeV the values differ 
much less.
The K$^+$ multiplicity per participating nucleon M(K$^+$)/$<A_{part}>$   
is similar for C+C and C+Au collisions at 1 AGeV and at 1.8 AGeV. 
These observations indicate that in C+Au collisions at 1 AGeV the pions 
are reabsorbed in the large gold nucleus (this effect is not expected
for K$^+$ mesons due to their antistrange quark content).      
At a beam energy of 1.8 AGeV, however, energetic secondary collisions 
of projectile nucleons with nucleons in the Au target may contribute
to pion production  thus compensating for losses due to reabsorption.

In contrast to the K$^+$ mesons, the K$^-$ multiplicity per 
participating nucleon decreases by a factor of 0.55 
in C+Au as compared to C+C collisions at 1.8 AGeV. 
This result indicates
that K$^-$ mesons are strongly absorbed by the heavy target nucleus.   
This effect would also explain that the  
K$^-$/K$^+$ ratios at equivalent energies are smaller for C+Au than for C+C
collisions.

RBUU transport model calculations predict the K$^+$ data  
within a factor of two depending on the assumption on the in-medium effective
mass of the K$^+$ mesons. 
For C+C collisions at 1  AGeV, the K$^+$ data favor the 
calculation based on the in-medium 
kaon mass whereas for C+C collisions at 1.8 AGeV
the bare mass option is closer to the data. For C+Au collisions
both assumtions on the K$^+$ effective in-medium mass explain the data 
equally well. 

The yield of K$^-$ mesons measured in C+C collisions at 1.8 AGeV 
is clearly underpredicted by the  transport calculations 
if a bare antikaon mass is assumed. The agreement with the data is improved 
when assuming  an attractive antikaon-nucleon
potential in the nuclear medium. For C+Au collisions at 1.8 AGeV the data
can hardly distinguish between the two options of the K$^-$ in-medium effective
mass. In this case the effect of the in-medium potential $-$ which depends
on the baryon density $-$ may be washed out by the   
the strong absorption of K$^-$ in the gold nucleus. 

In summary, the  RBUU transport calculations describe  
both kaon and antikaon data within a factor of about two
with consistent assumptions on in-medium effects. The accuracy 
of the calculations has to be further improved in order to extract  
conclusive informations on the in-medium properties of strange mesons.

\section{Acknowledgement}
We thank Elena Bratkowskaya and Wolfgang Cassing     
for providing us with the results of their transport calculations
and for various discussions. 
This work was supported by the German Federal Government (BMBF), by the
Polish Committee of Scientific Research (Contract No. 2P03B11515) and 
by the GSI fund for University collaborations.

\newpage 

\begin{table*}[H]
\caption{Beam energy, target isotope and thickness;
approximate rates and total numbers of identified kaons and antikaons.
The beam intensity is 10$^8$ ions 
per spill which had an average length of 4 s with
a pause of 2 s (at 1 AGeV) and 4 s at (1.8 AGeV). The pion recording rate 
is typically 1-2$\times$$10^5$/h.}
\label{table1} 
\begin{tabular}{lllllll}
\hline\noalign{\smallskip}
 beam & target &target &K$^+$&K$^-$&K$^+$&K$^-$\\ 
energy & &thickness &rate&rate&number&number\\
\noalign{\smallskip}\hline\noalign{\smallskip}
1.0 AGeV&$^{12}$C  & 5 mm & 25/h& $-$&1200& $-$\\
1.8 AGeV&$^{12}$C  & 5  mm & 900/h&25/h&17000&2000\\
1.0 AGeV&$^{197}$Au  & 0.5 mm &120/h& $-$&770& $-$\\
1.8 AGeV&$^{197}$Au  & 0.5 mm & 3000/h& 70/h& 13000&1300\\
\noalign{\smallskip}\hline
\end{tabular}
\end{table*}

\begin{table*}[H]
\caption{ System, beam energy, laboratory angles, inverse slope parameters, 
inclusive production cross sections and multiplicities per participating 
nucleon
for pions, kaons and antikaons. The values for T and $\sigma$ are determined 
by fitting a Boltzmann distribution d$^3\sigma$/dp$^3$ $\propto$ exp(-E/T)
to the data. In the case of pions a superposition of two 
Boltzmann distributions is fitted and the slope parameter of 
the high energy pions (T$_2$) quoted. 
In the extrapolation to full phase space 
the nonisotropic angular distribution is taken into account (see text).     
The definition of $M/<A_{part}>$ is given in the text. 
The quoted errors on the cross sections include systematic effects.
}
\begin{tabular}{lllllllll}
\hline\noalign{\smallskip}
system&beam energy & $\Theta_{lab}$ &T$_2$($\pi^+$)&T(K$^+$)&$\sigma$($\pi^+$)
&$\sigma$(K$^+$) 
& $\frac{M(\pi^+)}{<A_{part}>}$ & $\frac{M(K^+)}{<A_{part}>}$\\
&(AGeV) &  &(MeV) & (MeV) & (b)& (mb)& $\times10^{-2}$& $\times10^{-5}$\\
\noalign{\smallskip}\hline\noalign{\smallskip}
C+C& 1.0  & 44$^{\circ}$, 70$^{\circ}$ &57$\pm$5   & 58$\pm$6 &0.35$\pm$0.05 
&0.1$\pm$0.02 & 6.1$\pm$0.9 & 1.8$\pm$0.35 \\
C+Au& 1.0  & 44$^{\circ}$, 70$^{\circ}$ & 68$\pm$5  & 76$\pm$5 &1.2$\pm$0.2 
& 1.2$\pm$0.2 & 1.8$\pm$0.3 & 1.8$\pm$0.3 \\
C+C& 1.8  & 32$^{\circ}$, 40$^{\circ}$, 48$^{\circ}$, 60$^{\circ}$ & 76$\pm$5   
&75$\pm$5 &0.69$\pm$0.1 &3.0$\pm$0.3 &12.1$\pm$1.8 & 53$\pm$5\\
C+Au& 1.8  & 40$^{\circ}$, 60$^{\circ}$ & 80$\pm$6  & 90$\pm$6 &5.0$\pm$0.7 
& 30$\pm$5& 7.6$\pm$1 & 45$\pm$8 \\
\noalign{\smallskip}\hline
\noalign{\smallskip}\hline
system&beam energy & $\Theta_{lab}$ &T$_2$($\pi^-$)&T(K$^-$)&$\sigma$($\pi^-$)
&$\sigma$(K$^-$) 
&$\frac{M(\pi^-)}{<A_{part}>}$ & $\frac{M(K^-)}{<A_{part}>}$\\
&(AGeV) &  &(MeV) & (MeV) & (b)& (mb) & $\times10^{-2}$& $\times10^{-5}$\\
\noalign{\smallskip}\hline
C+C&1.8  & 40$^{\circ}$, 60$^{\circ}$ & 76$\pm$6 & 55$\pm$6  &0.64$\pm$0.1 
& 0.076$\pm$0.02 & 11.2$\pm$1.8 & 1.3$\pm$0.35\\
C+Au&1.8  & 40$^{\circ}$, 60$^{\circ}$ & 82$\pm$6 & 73$\pm$7  & 6.0$\pm$1
& 0.5$\pm$0.15  & 9.1$\pm$1.5 & 0.75$\pm$0.23\\
\noalign{\smallskip}\hline
\end{tabular}
\label{table2}
\end{table*}

\newpage

\vspace{-1.cm}

\vspace{-1.cm}

\begin{thebibliography}{50}
\bibitem{stock} R. Stock, Phys. Rep. {\bf 135} (1986) 259
\bibitem{cass_brat}  W. Cassing, E. Bratkovskaya, Phys. Rep. {\bf 308} 
(1999) 65
\bibitem{sen_str} P. Senger, H. Str\"obele,
J.Phys. G: Nucl. Part. Phys. {\bf 25} (1999) R59
\bibitem{harris} J. W. Harris et al., Phys. Lett. {\bf 153 B}  (1985) 377
\bibitem{aich_ko} J. Aichelin, C. M. Ko, Phys. Rev. Lett. {\bf 55} (1985) 2661
\bibitem{li_ko} G. Q. Li, C. M. Ko, Phys. Lett. {\bf B 349} (1995) 405
\bibitem{brown1} G.E. Brown, C.H. Lee, M. Rho, V. Thorsson,
Nucl. Phys. {\bf A 567} (1994) 937
\bibitem{waas} T. Waas, N. Kaiser, W. Weise, Phys. Lett. {\bf B 379} 
(1996) 34
\bibitem{schaffner} J. Schaffner-Bielich, J. Bondorf, I. Mishustin,
Nucl. Phys. {\bf A 625} (1997) 325
\bibitem{lutz}  M. Lutz, Phys. Lett. {\bf B 426} (1998) 12
\bibitem{brobet} G.E. Brown, H.A. Bethe, Astrophys. Jour. {\bf 423} 
(1994) 659
\bibitem{li_lee_br} G.Q. Li, C.H. Lee, G.E. Brown,
Phys. Rev. Lett. {\bf 79} (1997) 5214
\bibitem{barth} R. Barth et al., Phys. Rev. Lett. {\bf 78} (1997) 4007
\bibitem{laue} F. Laue et al., Phys. Rev. Lett. {\bf 82} (1999) 1640 
\bibitem{cassing} W. Cassing et al., Nucl. Phys. {\bf A 614} (1997) 415
\bibitem{li_ko_fang} G. Q. Li, C. M. Ko, X. S. Fang, Phys. Lett. {\bf B 329} (1994) 149
\bibitem{shin} Y. Shin et al., Phys. Rev. Lett. {\bf 81} (1998) 1576
\bibitem{li_ko_br}  G. Q. Li, C. M. Ko and G. E. Brown, Phys. Lett. 
{\bf B 381} (1996) 17
\bibitem{ritman} J. Ritman et al., Z. Phys. {\bf A 352}  (1995) 355
\bibitem{li_flow} G. Q. Li et al., Phys. Rev. Lett. {\bf 74} (1995) 235
and Phys.Lett. {\bf B 381} (1996) 17
\bibitem{pbm1} P. Braun-Munzinger, J. Stachel, J.P. Wessels, N. Xu,
Phys. Lett. {\bf B 344} (1995) 43
\bibitem{cleymans} J. Cleymans, H. Oeschler, K. Redlich,
Phys. Rev. {\bf C 59} (1999) 1663
\bibitem{schnetzer} S. Schnetzer et al., Phys. Rev. Lett. {\bf 49 } 
(1982) 989 and Phys.Rev. {\bf C 40 } (1989) 640
\bibitem{shor} A. Shor et al., Phys. Rev. Lett. {\bf 63} (1989) 2192
\bibitem{schroeter} A. Schr\"oter et al., Z. Phys. {\bf A 350} (1994) 101
\bibitem{senger} P. Senger et al., Nucl. Instr. Meth. {\bf A 327} (1993) 393
\bibitem{misko2} D. Mi\'skowiec et al., 
Nucl. Instr. Meth. {\bf A 350} (1994) 174 
\bibitem{huefner} J. H\"ufner, J. Knoll, Nucl. Phys. {\bf A 290} (1977) 460
\bibitem{balewski} J. Balewski et al., Phys. Lett. {\bf B 388} (1996) 859
and Phys. Lett. {\bf B 420} (1998) 211 
\bibitem{averbeck} R. Averbeck et al., Z. Phys. {\bf A 359} (1977) 215
\bibitem{dover} C. B. Dover, G. E. Walker, Phys. Rep. {\bf 89} (1982) 1
\bibitem{si_ca_ko} A. Sibirtsev, W. Cassing and C.M. Ko, Z. Phys.
{\bf A 358} (1997) 101
\bibitem{brat_cass} E. Bratkovskaya, W. Cassing, U. Mosel, Nucl. Phys. 
{\bf A 622} (1997) 593
\bibitem{sibirtsev} A. Sibirtsev, Phys. Lett. {\bf B 359}  (1995) 29
\bibitem{balestra} F. Balestra et al., Phys. Lett. {\bf B 468} (1999) 7
\bibitem{fuchs}  C. Fuchs et al., Phys. Rev. {\bf C 56} (1997) R606
\bibitem{ai_fu} C. Fuchs, J. Aichelin, private communication
\bibitem{brat_priv} E. Bratkovskaya, W. Cassing, private communication


\end{thebibliography}
\end{document}